\RequirePackage{fix-cm}
\documentclass{svjour3}                     
\smartqed  
\usepackage{graphicx}
\usepackage{amsmath,amsfonts}
\usepackage{algorithmic}
\usepackage{algorithm}
\usepackage{array}
\usepackage{adjustbox}
\usepackage{subcaption}
\usepackage[normalem]{ulem}
\usepackage{rotating}
\usepackage{booktabs}
\useunder{\uline}{\ul}{}
\usepackage{textcomp}
\usepackage{stfloats}
\usepackage{url}
\usepackage{threeparttable}
\usepackage{makecell}
\usepackage{multirow}
\usepackage{verbatim}
\usepackage{graphicx}
\usepackage{float}
\usepackage{natbib}
\usepackage{xurl}
\usepackage[
colorlinks=true,
citecolor=blue,
linkcolor=blue,
urlcolor=black
]{hyperref}

\newcommand{\Coder}{\textit{SynthCoder}}

\newcommand{\CoderSeed}{\textit{SynthCoder-Seed}}

\newcommand{\CoderQwen}{\textit{SynthCoder-Qwen}}

\newcommand{\SC}{\textit{SC}}
\newcommand{\SCSeed}{\textit{SC-Seed}}
\newcommand{\SCQwen}{\textit{SC-Qwen}}

\newcommand{\promptbar}{|}
\newcommand{\promptscore}{\_}

\usepackage{xcolor}
\definecolor{DarkRed}{RGB}{139,0,0}
\definecolor{DarkBlue}{RGB}{0,0,139}
\definecolor{DeepOrange}{RGB}{204,85,0}
\definecolor{DeepGreen}{RGB}{0,100,0}
\newcommand{\DO}[1]{\textcolor{DeepOrange}{\textit{#1}}}
\newcommand{\DG}[1]{\textcolor{DeepGreen}{\textit{#1}}}
\newcommand{\DR}[1]{\textcolor{DarkRed}{\textit{#1}}}

\newcommand{\yellowBG}[1]{
    \begingroup
    \setlength{\fboxsep}{1pt}
    \colorbox{yellow}{\vphantom{Ag}\smash{\texttt{#1}}}
    \endgroup
}
\newcommand{\orangeBG}[1]{
    \begingroup
    \setlength{\fboxsep}{1pt}
    \colorbox{orange}{\color{white}\vphantom{Ag}\smash{\texttt{#1}}}
    \endgroup
}
\newcommand{\redBG}[1]{
    \begingroup
    \setlength{\fboxsep}{1pt}
    \colorbox[HTML]{F8CECC}{\vphantom{Ag}\smash{\texttt{#1}}}
    \endgroup
}

\begin{document}

\title{{\Coder}:
Anti-Pattern Identification and Model Training for FIM Mode Code Completion
}

\titlerunning{{\Coder}}

\author{Dongjun~Yu \and
Xiao~Yan \and
Zhenrui~Li \and
Jipeng~Xiao \and
Haochuan~He \and
Yongda~Yu \and
Hao~Zhang \and
Guoping~Rong\textsuperscript{*} \and
Xiaobo~Huang
}

\institute{
Dongjun~Yu,
Xiao~Yan,
Zhenrui~Li,
Jipeng~Xiao,
Haochuan~He,
Yongda~Yu,
Hao~Zhang  \\ and
Guoping~Rong \at
Software Institute,
Nanjing University, Nanjing, China. \\
\email{\{
502023320014,
yanxiao,
522024320090,
231250096,
231250169,
yuyongda, \\
hao-zhang
\}@smail.nju.edu.cn,
ronggp@nju.edu.cn} \\
\textsuperscript{*} Guoping~Rong is the \textit{corresponding author}.
\and
Xiaobo~Huang \at
Department of Computer Science, Beijing Normal-Hong Kong Baptist University, Zhuhai,
China. \\
\email{shengjibanzi@qq.com}
}

\date{Received: date / Accepted: date}

\maketitle

\begin{abstract}

As a leading application of large language models (LLMs) in software engineering, Fill-in-the-Middle (FIM) mode code completion has drawn wide attention. Training such models requires masking code corpora, yet common strategies tend to be problematic. For example, character-delimited random masking may yield many unrealistic cases (e.g., cutting keywords or identifiers),
while purely AST-based masking cannot mask concurrent elements (subtrees) that span across multiple subtrees. These limitations easily create anti-patterns that rarely occur in real-world code completion scenarios, diminishing FIM performance.
We introduce \Coder, which adopts optimized masking strategies better aligned with developer's expectation in FIM. Specifically, we first refine AST-level node masking and add heuristics that better mimic developers' expectation to construct the training corpora.
Subsequently, \CoderSeed\ and \CoderQwen, built upon Seed-Coder-8B-Base and Qwen2.5-Coder-7B respectively, employ a two-stage training pipeline, i.e., curriculum-based fine-tuning stage followed by Direct Preference Optimization (DPO) alignment stage with rejected code sampled preference data. Besides, to suppress erroneous context repetition, we include negative samples that duplicate existing code during DPO, mitigating code-echo failures where the model copies neighboring context instead of producing a valid completion.
Extensive experiments on Santacoder-fim-task, aiXcoder-FIM-Evaluation and CrossCodeEval benchmarks show that the models trained with our mitigation strategies improve over mainstream baselines on Exact Match (EM) and Edit Similarity (ES) for text-based FIM benchmarks, and on Pass@1 for Santacoder-fim-task, a benchmark with test cases. \Coder\ also yields less code-echo and consumes fewer tokens at inference, yielding higher practical efficiency. Ablation studies further support the contribution of our optimized masking and repetition-suppression mechanisms.

\keywords{LLM \and Code completion \and FIM \and Post-training }
\end{abstract}

\section{Introduction}\label{section:introduction}

{C}{ode} completion accelerates software development by predicting a developer's coding intentions and providing syntax hints and code snippets.
In recent years, Large Language Models (LLMs) have brought transformative changes with impressive capabilities in text understanding, reasoning, and generation~\citep{hou2024large}.
Products based on LLMs have emerged. Examples include the GitHub Copilot~\footnote{\url{https://github.com/features/copilot}} plugin, which is compatible with various IDEs, and the Cursor~\footnote{\url{https://www.cursor.com}} IDE, which integrates LLMs into various features of the development tool.
However, due to concerns such as information security, enterprises often choose to customize open-source models for fill-in-the-middle (FIM) code completion tasks, positioning FIM as a parallel application scenario alongside agent-based code generation.
Given response time requirements, most such models employ LLMs with fewer than 10B parameters~\citep{li2025aixcoder, jiang2024aixcoder, seed2025seedcoder,roziere2023code, hui2024qwen2, li2023starcoder}, making sub-10B FIM a cost-effective solution with adaptable deployment characteristics~\citep{Subramanian2025SmallSLMs,Wang2025ComprehensiveSLM}.

To further enhance compact code LLMs for FIM code completion, practitioners often apply various post-training techniques using customized datasets. Among these, masking complete code segments is a common strategy for preparing training corpora. Currently, two primary approaches prevail: the first relies exclusively on character-level random masking~such as CodeLlama~\citep{roziere2023code} , InCoder~\citep{fried2022incoder},
and Seed-Coder series~\citep{seed2025seedcoder}, while the second uses Abstract Syntax Trees (ASTs) to guide masking over syntactic elements, as seen in the direct AST-node random masking practice used by the Qwen2.5-Coder series~\citep{hui2024qwen2}.

\begin{figure}[h]
    \centering
    \begin{subfigure}[t]{0.49\textwidth}
        \centering
        \includegraphics[width=\linewidth]{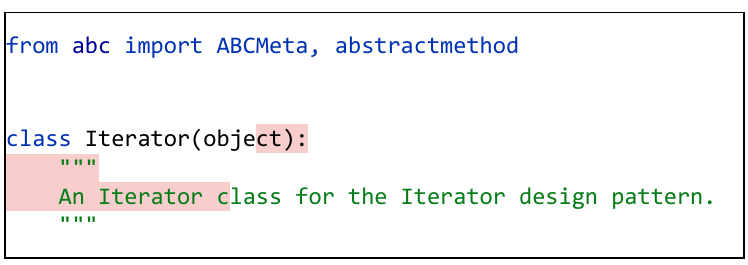}
        \caption{Truncated Identifiers and AST Units (the Mask Part Is \redBG{red}, Resulting in `\texttt{obje}' and `\texttt{lass}').}
        \label{fig:anti-pattern-a}
    \end{subfigure}
    \hfill
    \begin{subfigure}[t]{0.49\textwidth}
        \centering
        \includegraphics[width=\linewidth]{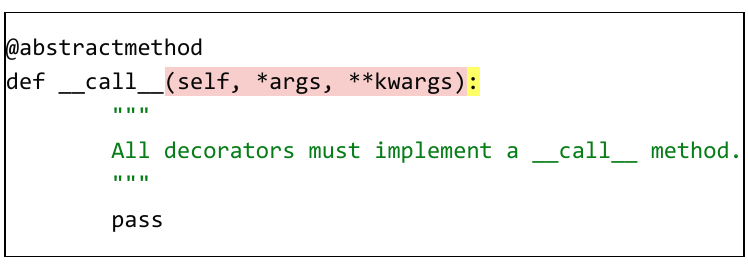}
        \caption{The \redBG{red} Part Will Be Masked in Regular AST-Based Masking, Yet the Trailing Part (Colored \yellowBG{yellow}) Should Also Be Masked to Better Reflect Developers' Expectation on FIM Tasks.}
        \label{fig:anti-pattern-b}
    \end{subfigure}
    \caption{Anti-Pattern Examples in ExecRepoBench~\citep{yang2024execrepobench}.}
    \label{fig:anti-pattern}
\end{figure}

However, these masking approaches may suffer from notable limitations. On one hand, random masking based on arbitrary delimiters can generate a large number of unrealistic scenarios rarely encountered in real-world coding practice~\citep{roziere2023code, yang2024execrepobench}. As illustrated in \textbf{Figure~\ref{fig:anti-pattern-a}}, masked spans may begin and/or end with truncated identifiers.
For human developers performing code completion, such a  scenario -- where both the beginning and the end consist of syntactically incomplete elements -- is highly bizarre.
On the other hand, AST-based random masking, while syntax-aware, struggles to capture some types of frequent contexts that arise in real development~\citep{yang2024execrepobench}. For example, as is shown in \textbf{Figure~\ref{fig:anti-pattern-b}}, when defining a function, developers often expect to complete the entire line after the function name, which requires the masking part spanning across multiple AST subtrees. However, normally an AST-based masking strategy can only capture the red subtree corresponding to the parameter part, while missing the yellow portion such as the trailing symbol `\texttt{:}', failing to reflect such common code completion expectations. Similar phenomena are frequently observed with the aforementioned masking strategies. They can reduce the realism and consistency of training instances, which may in turn affect the FIM performance of the model trained on the dataset.

To address these issues, we introduce the \Coder\ series, optimized for code completion through human-expectation-aligned optimized masking strategies. By applying a two-stage post-training pipeline including both Supervised Fine-Tuning (SFT) and Direct Preference Optimization (DPO)~\citep{rafailov2023direct}, we trained Seed-Coder-8B-Base and Qwen2.5-Coder-7B to yield \CoderSeed and \CoderQwen, respectively.

Specifically, we first engaged experienced developers to systematically identify phenomena in the corpora -- constructed using both random masking and AST-based masking strategies -- that deviate from the common expectations of human developers. These deviations are referred to as anti-patterns. For each anti-pattern, we designed corresponding resolution strategies. Following these new strategies, we reconstructed the masked training corpora and incorporated additional negative examples. We selected Seed-Coder-8B-Base~\citep{seed2025seedcoder} and Qwen2.5-Coder-7B~\citep{hui2024qwen2}, two recent compact SLM-scale FIM-capable code base models from different model families and with different native FIM prompt conventions. Multiple training approaches, including curriculum learning~\citep{sagtani2025improving} and DPO~\citep{rafailov2023direct}, were then applied to these two models to enhance model's FIM performance.

Extensive experiments on three public datasets, i.e.
Santacoder-fim-task~\citep{allal2023santacoderdontreachstars}, aiXcoder-FIM-Evaluation~\citep{jiang2024aixcoder} and CrossCodeEval ~\citep{ding2023crosscodeeval} show that \Coder\ improves over mainstream baselines -- including models such as aiXcoder~\citep{jiang2024aixcoder, li2025aixcoder}, StarCoder~\citep{li2023starcoder}, DeepSeek~\citep{guo2024deepseekcoder}, Qwen~\citep{hui2024qwen2}, and Seed~\citep{seed2025seedcoder} series -- on Exact Match (EM) and Edit Similarity (ES) for text-based FIM benchmarks, and on Pass@1 for Santacoder-fim-task, a benchmark with test cases. \Coder\ also produces less code-echo and consumes fewer tokens at inference, indicating higher practical efficiency.
Ablation studies confirmed  the importance of the optimized masking strategies and repetition-suppression mechanisms.

The main contributions of this paper are as follows:
\begin{enumerate}
    \item From a pragmatic perspective, we systematically identified various anti-patterns in FIM training-data construction that may reduce the realism of completion scenarios. Some align with prior research, such as \textbf{AP1}~\citep{gong2025structure}, and \textbf{AP2–AP4}~\citep{jiang2024aixcoder}, while others represent novel findings, such as \textbf{AP5} and \textbf{AP6} (cf. \textbf{Section~\ref{section:anti-pattern-analysis}}).

    \item For each anti-pattern, we designed corresponding mitigation or resolution strategies. These were implemented during training dataset construction and training process, resulting in \Coder\ -- post-trained variants of these compact code base models that achieve the best overall results among the evaluated models in the reported FIM evaluation settings.

    \item We conducted an empirical evaluation of \Coder\ on public datasets and performed ablation studies to thoroughly validate the effectiveness of the proposed mechanisms and methods on both Seed-Coder-8B-Base and Qwen2.5-Coder-7B for addressing anti-patterns in FIM model training.
\end{enumerate}

The rest of this paper is structured as follows.
\textbf{Section~\ref{section:related_work}} reviews related work. \textbf{Section~\ref{section:methodology}} provides a complete description of our entire method, from raw data retrieval, dataset construction to both of the model post-training stages.
To validate the effectiveness of our proposed method, in \textbf{Section~\ref{section:evaluation}}, we conduct comprehensive evaluations on \Coder\ with FIM benchmarks, followed by an in-depth analysis of the results in \textbf{Section~\ref{section:results}}. Based on these empirical findings, \textbf{Section~\ref{section:discussion}} discusses our considerations on FIM model training and application.
\textbf{Section~\ref{section:validity}} addresses the limitations of the current work. Finally, this paper concludes in \textbf{Section~\ref{section:conclusion}} with suggestions on future work.

\section{Related Work}\label{section:related_work}

This section describes the related studies to position our work.

\subsection{Training LLMs for Coding}\label{section:training-llms-for-coding}

Code generation and completion have long been fundamental capabilities of LLMs, as most foundation models incorporate vast amounts of source code in their initial pre-training data~\citep{jiang2024survey}. Post-training refers to the collection of techniques employed after the initial pre-training to refine, adapt, and align a model for specific tasks and user requirements, which has become a critical phase for specializing LLM for code-related tasks like code completion, code generation and other tasks~\citep{kumar2025llm}. To further enhance these capabilities and adapt models for specific programming scenarios, researchers have developed several distinct post-training approaches.

Initial efforts centered on SFT with carefully curated instruction-code datasets, as is demonstrated by Code Alpaca~\citep{wang2022self} which generated instruction-output pairs from seed tasks using ChatGPT, but its data diversity was limited to some extent. To address this, subsequent research proposed innovative methods: Evol-Instruct employed in WizardCoder~\citep{xu2023wizardlm} evolutionarily rewrites simple instructions, while OSS-Instruct~\citep{wei2023magicoder} innovatively uses open-source code as seeds to prompt an LLM to generate relevant pairs of problems and solutions.
In the instruction-tuning phase, Seed-Coder~\citep{seed2025seedcoder} generates a large-scale synthetic dataset by prompting an LLM with diverse, high-quality code and text sources. The resulting instruction-response pairs are then refined through a multi-stage process, including rule- and model-based filtering for quality and a final sandbox-based self-correction mechanism to verify and improve complex solutions.

To further tune model behavior towards subjective qualities like readability, alignment techniques are commonly used. While OpenAI Reinforcement Learning from Human Feedback (RLHF)~\citep{ouyang2022training} was a pioneering approach, recent work may prefer lightweight approaches like DPO~\citep{rafailov2023direct}. One typical method is the CodeDPO~\citep{zhang2024codedpo}, which constructs preference data through a self-verification mechanism, optimizing both the correctness and execution efficiency of the generated code.
More recent alignment algorithms, such as Group-Relative Policy Optimization (GRPO) used in DeepSeekMath~\citep{shao2024deepseekmath}, enable more fine-grained alignment without requiring a complex Reinforcement Learning (RL) pipeline and can also strengthen reasoning abilities. Initially, Chain of Thought (CoT) prompting~\citep{wei2022chain} improved code problem-solving by having models generate intermediate steps like plans~\citep{sun2024enhancing}. This has evolved into the development of specialized reasoning models like DeepSeek-R1~\citep{guo2025deepseekr1}, which leverage Inference Time Scaling~\citep{wu2024inference} to obtain stronger abilities in code and math. Reasoning variants of Seed-Coder~\citep{seed2025seedcoder} are also followed by a GRPO-based RL phase that utilizes a progressive exploration strategy to methodically improve multi-step reasoning for difficult tasks based on a limited set of complex, distilled coding problems to encourage diverse thinking patterns.
Reasoning models are becoming indispensable for other software engineering tasks such as code generation~\citep{chen2025r1} and reasoning~\citep{tang2025codereasoner}, though may be constrained in latency sensitive tasks like code completion~\citep{izadi2022codefill}.
Reinforcement learning can also leverage verifiable feedback from the programming environment. For example, KodCode~\citep{xu2025kodcode} incorporates the pass/fail outcomes of unit tests as a direct signal for improvement, either as a reward or as a filter for high-quality data.

\subsection{FIM Code Completion}\label{section:fim}

The FIM paradigm, proposed by OpenAI~\citep{bavarian2022efficient}, established a specialized input-output format for code completion that has since been widely adopted. Early works such as CodeLlama and InCoder, laid a foundational basis for this area~\citep{roziere2023code, fried2022incoder} and relied on character-level random masking to construct their training corpora. However, constructing high-quality training data remains a challenge, particularly in ensuring that the completion target is meaningful in relation to its surrounding context (the prefix and suffix) and covers complex and diverse scenarios in real-world development.

Recent work has explored syntax-aware FIM pretraining and evaluation. For instance, \citet{ast-t5} proposed AST-T5, which incorporates AST structure into pretraining without architectural changes, showing strong performance in code-to-code tasks. \citet{safim} introduced SAFIM, a syntax-aware FIM benchmark that evaluates LLMs on structured code completions, revealing that pretraining methods and data quality often outweigh model size. \citet{gong2025structure} extended this line by proposing structure-aware FIM pretraining with syntax-tree-based masking, though their method focuses on single or sibling AST nodes.
Similarly leveraging code's structural information, the ExecRepoBench team introduced a multi-level code completion method that also relies on ASTs to build training data. Their approach emphasizes enhancing the model's capabilities in practical coding scenarios that involve complex, multi-file interdependencies~\citep{yang2024execrepobench}. This reflects a trend in code completion research, which is progressively moving from the single-file level to more complex repository-level scenarios that include cross-file dependencies~\citep{yang2024execrepobench, ding2023crosscodeeval,zhang2023repocoder,wu2024repoformer}.

Researchers have already realized that common masking strategies to prepare the corpora for FIM model training may contain shortcomings.
For example,~\cite{roziere2023code} found that the character-delimited random masking method may generate subtokens that do not exist in the vocabulary.~\cite{jiang2024aixcoder} believes that the character-delimited random masking method may suggest an incomplete code snippet, which may deviate from developers' actual expectations.
Nevertheless, a systematic investigation of anti-patterns during FIM model training is still lacking, which motivates us to carry out this study.

In summary, OpenAI first proposed FIM paradigm~\citep{bavarian2022efficient}, which has since been widely adopted by code models. To construct training data for this paradigm, researchers have employed (1) character-level random masking and (2) AST-based random masking.
Prior studies have partially recognized the limitations of these masking strategies, as they often fail to meet developers' real expectations during coding. Yet a systematic investigation of such anti-patterns is still lacking, which motivates our study.

\section{Methodology}\label{section:methodology}

The details in anti-pattern identification and corresponding resolutions as well as details in \Coder\ model post-training stages will be elaborated in this section.

\subsection{Overview}
\begin{figure}[htb]
    \centering
    \includegraphics[width=0.87\textwidth]{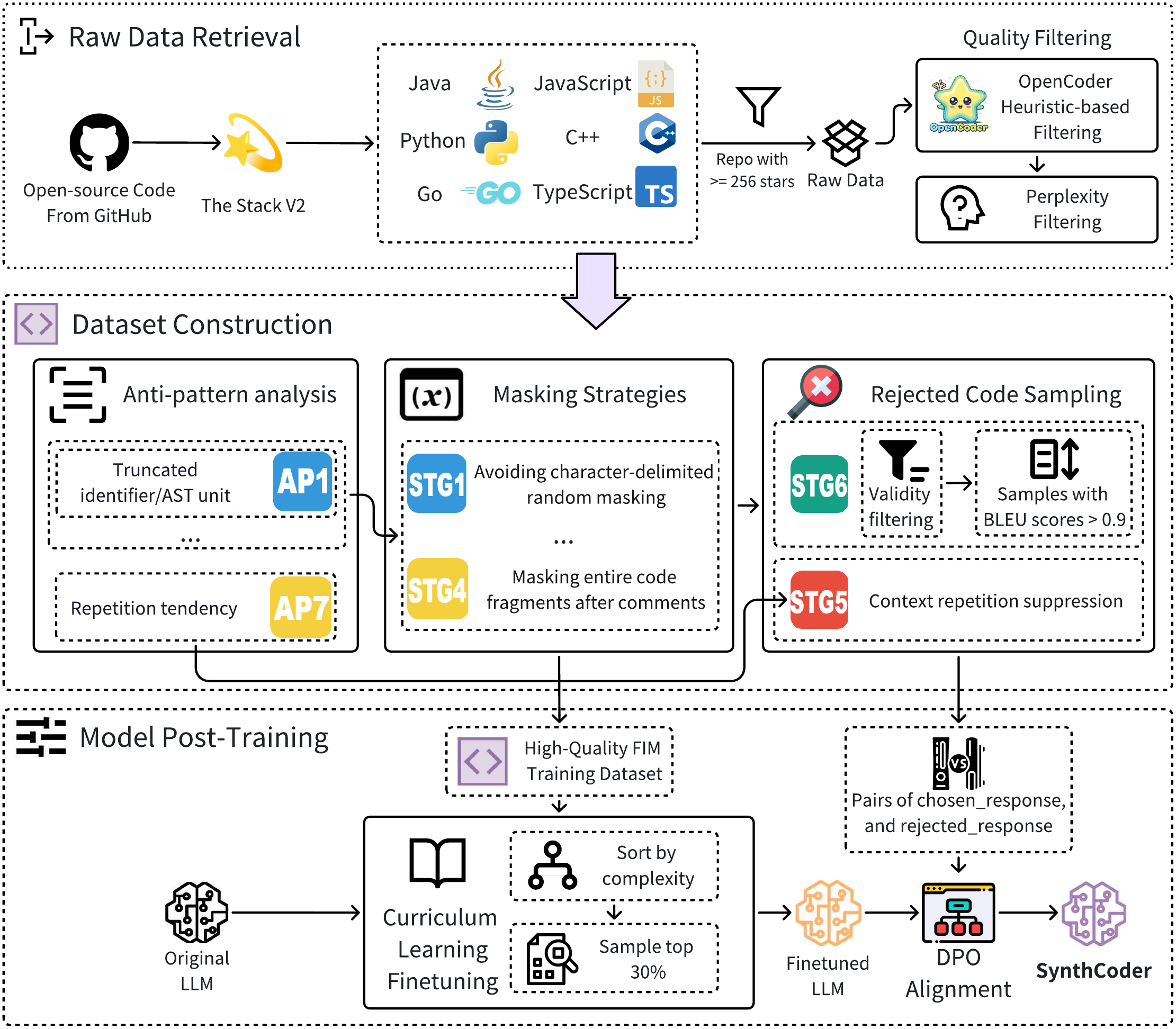}
    \caption{The Overall Methodology of \Coder.}
    \label{fig:methodology-framework}
\end{figure}

\textbf{Figure~\ref{fig:methodology-framework}} provides an overview of our overall methodology. In general, our approach consists of three main components: raw data retrieval, dataset construction, and model post-training. The primary focus lies on the dataset construction phase, in which we apply a systematic method to identify anti-patterns and develop targeted strategies to address them, thereby facilitating the development of our training corpora. Further details of the methodology are provided in the subsequent sections.

\subsection{Raw Data Retrieval}\label{section:data-sources}

We started with the officially deduplicated subset~\footnote{\url{https://huggingface.co/datasets/bigcode/the-stack-v2-train-smol-ids}} of the Stack v2~\citep{lozhkov2024starcoder} dataset, which is one of the biggest code corpora.
To balance data quality and acquisition cost, we curated a specialized subset by selecting files from repositories with 256 or more stars,
focusing on six popular languages, i.e., Java (25\%), Python (25\%), Go (14\%), JavaScript (14\%), C++ (14\%), and TypeScript (8\%). This final collection served as the basis for our FIM training set construction. We then applied two successive filtering stages to the raw data: first a heuristic rule-based pipeline for corpus cleaning, followed by model-specific perplexity-based sample selection for constructing the post-training corpora.
Before constructing FIM instances, we also applied a repository-level leakage-control step. For benchmarks adopted by subsequent evaluations with identifiable source repositories, we collected their repository identifiers and removed the entire corresponding repositories from the corpora. Thus, if a benchmark contains any sample from a repository, no file from that repository is used in the subsequent masking, filtering, SFT, or DPO data construction stage.

\subsubsection{Heuristics-Based Filtering}

OpenCoder's \texttt{opc\_data\_filtering}~\footnote{\url{https://github.com/OpenCoder-llm/opc\_data\_filtering}} is a heuristic filtering framework for large-scale code pre-training corpora, which considers the unique characteristics of different programming languages and includes over 100 filtering rules~\citep{huang2024opencoder}, with guiding principle to remove harmful data while ensuring that the overall distribution of the dataset is not significantly affected.
In this work, we also adopt this pipeline to filter the raw code data based on the built-in rules in the pipeline, thus obtaining a code dataset with relatively high quality yet remaining diversity.

\subsubsection{Perplexity-Based Filtering}

Perplexity (PPL) measures the certainty of a model's predictions for a given sample. A model that outputs a low perplexity for a sample is better at predicting it~\citep{chen1999empirical}. For a sequence $x=(x_1, x_2,...,x_N)$ of length $N$, where the model's predicted conditional probability is $P(x_t | x_{<t})$, the perplexity $\text{PPL}(x)$ is as follows:

\begin{equation*}
\text{PPL}(x) = \exp\Big(-\frac{1}{N}\sum_{t=1}^{N}\log P(x_t|x_{<t})\Big)
\end{equation*}

In this work, unless otherwise specified, PPL scoring is performed with the base model corresponding to the target post-training family. Since perplexity is model-dependent, the same sample can receive different PPL scores under different base models; the filtered corpus is therefore constructed separately for each model family.
For the post-training process, data with very low perplexity represents knowledge the LLM has already mastered and may generate outputs correctly without further training. Conversely, data with very high perplexity deviates significantly from the distribution of the LLM's base data, and this subset tends to have a higher proportion of erroneous data. Accordingly, we discard the lowest and highest 5\% of the completion data by perplexity to construct a base-model-specific training corpus and improve the overall quality of the training set~\citep{ankner2024perplexed, wu2025mitigating}.

\subsection{Dataset Construction}\label{section:data-construction}

The core objective of our dataset construction is to simulate, to the greatest extent possible, the coding practices, contextual needs, and code completion expectations of human engineers in real-world programming tasks. This section elaborates the process to construct such a training corpora to meet this objective.

\subsubsection{Anti-pattern Analysis} \label{section:anti-pattern-analysis}

We employed Thematic Analysis~\citep{cruzes2011recommended} as the major qualitative research method to identify, analyze, and report patterns of bizarre codes after masking (we use term anti-pattern to describe these bizarre patterns) embedded within data samples. In this qualitative stage, the masked candidates were generated with the two mainstream FIM masking-rule families reviewed in \textbf{Section~\ref{section:introduction}} and \textbf{Section~\ref{section:fim}}: delimiter-based random masking and AST-based random masking. For the AST-based side, we focus on the direct practice of randomly selecting syntax-tree nodes as the masked span, as exemplified by Qwen2.5-Coder~\citep{hui2024qwen2}. Thus, the masking rules used for thematic analysis are the object of diagnosis: we first identify the anti-patterns produced by these existing rules, and then propose optimized masking strategies to mitigate them during training-corpus construction. Details of the Thematic Analysis process are elaborated as below.

\paragraph*{\uline{Background of team members.}} \
The research team consists of three software engineers with over 5 years of Python and Java programming experience, one of whom comes from the industry with over 15 years of programming experience in both languages, ensuring high credibility in the analysis process and analysis results as well.

\paragraph*{\uline{Iterative analysis workflow.}} \
The primary goal of the team was to examine each masked code snippet and, based on common programming conventions and their experience, determine whether the completion scenario aligned with typical programming practices of human developers; otherwise, it was classified as an anti-pattern. Meanwhile, the team also needed to categorize and label any anti-patterns exhibited by the masked code. A fundamental rule was established: all conclusions required consensus from at least two members with no opposing objections. Otherwise, a team discussion was required to reach consensus.

The workflow consisted of the following steps:

\begin{enumerate}
    \item \textbf{Data Collection:} We began by randomly sampling code snippets from the filtered Stack V2 dataset mentioned earlier, selecting 1,000 samples for each combination of programming language and masking strategy. This produced 4,000 masked candidates in total: Java and Python each contributed 2,000 candidates, split evenly between delimiter-based random masking and AST-based random masking, so each language-by-strategy cell contained 1,000 masked candidates.

    \item \textbf{Anti-pattern Identification:} One researcher with stronger expertise in Java was assigned to review the 2,000 masked Java candidates, while another more proficient in Python examined the 2,000 masked Python candidates, with the goal of identifying potential anti-patterns. The final classification of code segments as anti-patterns was then determined through a team-based discussion round.

    \item \textbf{Labeling:} We adopted an iterative process to assign labels to each instance of anti-pattern code snippets. The guiding principle was to prioritize the use of existing labels to characterize an anti-pattern wherever applicable. Only when no existing label adequately captured its features was a new label created. This process continued iteratively until all identified anti-pattern code snippets had been reviewed and tagged.
\end{enumerate}

Fleiss' $\kappa$ is applied to determine the degree of agreement among the three raters, achieving $\kappa=0.82$ and indicating high inter-rater reliability~\citep{ElEmam1999BenchmarkingKappa,GonzalezPrieto2023ReliabilityInterCoderAgreement}.
Throughout the Thematic Analysis, the team identified six typical anti-patterns (APs) within the samples. Several of these, such as \textbf{AP1} ~\citep{gong2025structure} and \textbf{AP2–AP4} ~\citep{jiang2024aixcoder}, were also observed in other open-source code completion training datasets. Additionally, we identified novel anti-patterns, namely \textbf{AP5} and \textbf{AP6}. The issues are analyzed as follows:

\begin{figure*}[htbp]
    \centering
    \begin{subfigure}[t]{0.49\textwidth}
        \centering
        \includegraphics[width=\linewidth]{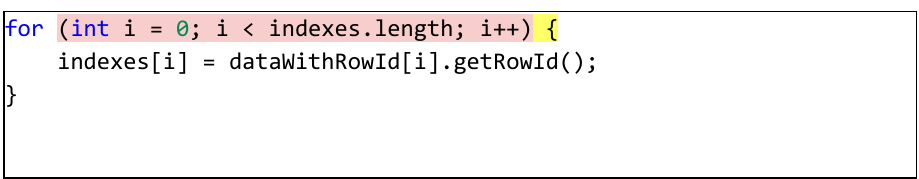}
        \caption{Incomplete Single-Line Completion Starting from `\texttt{for}' (Highlighted Mask).}
        \label{fig:sft-heuristic-first}
    \end{subfigure}
    \hfill
    \begin{subfigure}[t]{0.49\textwidth}
        \centering
        \includegraphics[width=\linewidth]{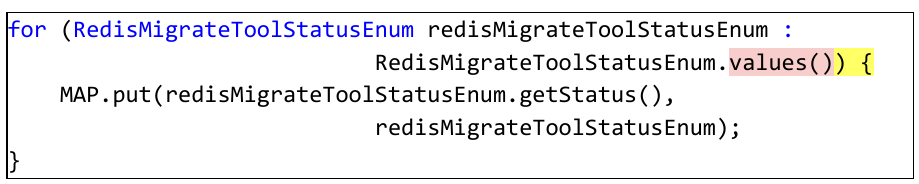}
        \caption{Incomplete Single-Line Completion Starting from `\texttt{.}' (Highlighted Mask).}
        \label{fig:sft-heuristic-second}
    \end{subfigure}

    \vspace{10 pt}
    \begin{subfigure}[t]{0.49\textwidth}
        \centering
        \includegraphics[width=\linewidth]{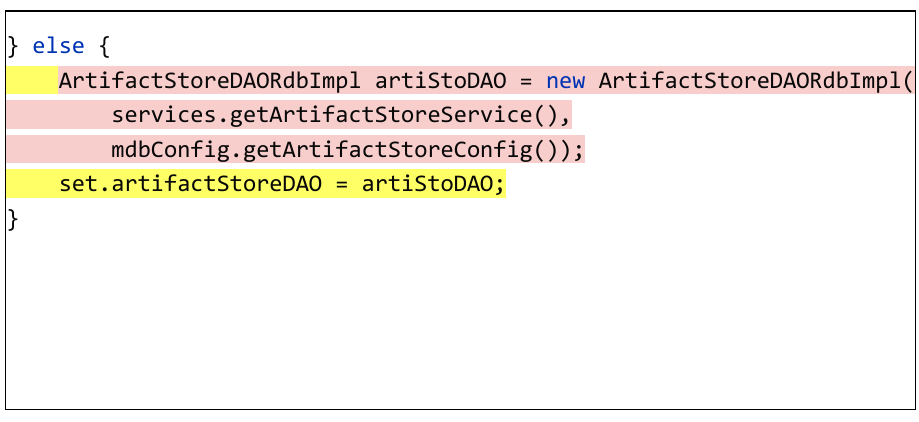}
        \caption{Bracket-Based Code Completion (Highlighted Mask).}
        \label{fig:sft-heuristic-third}
    \end{subfigure}
    \hfill
    \begin{subfigure}[t]{0.49\textwidth}
        \centering
        \includegraphics[width=\linewidth]
        {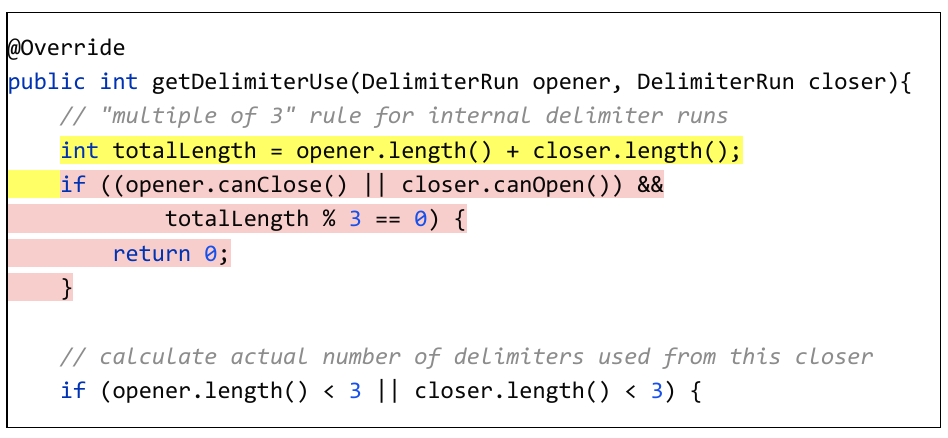}
        \caption{Comment-Following Code Completion (Highlighted Mask).}
        \label{fig:sft-heuristic-fourth}
    \end{subfigure}

    \raggedright

    {
    \vspace{0.5 em}
     \textsuperscript{\dag} The \redBG{red region} represents the actual AST-masked span; the \yellowBG{yellow region} together with the \redBG{red region} illustrates the regular code completion intentions (both should be masked).
    }

    \caption{Heuristic Dataset Construction Strategy Simulating Developer Behavior.}
    \label{fig:sft-heuristic}
\end{figure*}

\begin{itemize}

\item \textbf{AP1: Truncated identifiers and AST units.} Delimiter-based random masking often results in truncated identifiers or other incomplete minimal AST units (\textbf{Figure~\ref{fig:anti-pattern-a}}, partial comment fragments) due to random cutting at arbitrary boundaries. In real-world coding practice, developers generally will not encounter and do not expect such partial completions.

\item \textbf{AP2: Masked parameter AST nodes missing trailing content.} When a parameter AST node is masked (\textbf{Figure~\ref{fig:anti-pattern-b}}, red part in the figure), content at the end of the line that does not belong to the same AST node is omitted (yellow part, `:'). In practice, developers prefer to complete the entire line after entering the function name, rather than only masking the parameter node while leaving the trailing symbol unmasked.

\item \textbf{AP3: Masked condition AST nodes missing trailing content.} In a for statement, the loop condition AST node is masked (red part in the \textbf{Figure~\ref{fig:sft-heuristic-first}}), but trailing content at the end of the line outside the same AST node is omitted (yellow part, `\{’). Developers may expect the completion to fill the entire line after entering keywords like \emph{for}.

\item \textbf{AP4: Masked expression or call AST nodes missing trailing content.} For function call expressions following a symbol such as `.', the corresponding AST node is masked (red part in the \textbf{Figure~\ref{fig:sft-heuristic-second}}), but trailing content at the end of the line outside this node is omitted (yellow part). Developers generally expect the code completion to fill the line in such cases.

\item \textbf{AP5: Masked parenthesized content not preserving semantic integrity.} Content inside parentheses often forms a single semantic unit. When developers type paired parentheses, they expect the completion to fill the entire content. However, when the content spans multiple parallel AST nodes, AST-based masking can only cover one node (red part in the \textbf{Figure~\ref{fig:sft-heuristic-third}}), leaving other parts (yellow part) unmasked.

\item \textbf{AP6: Masked content following comments not preserving semantic integrity.} Complete single-line or multi-line comments often describe the functionality of subsequent code blocks. When the model cannot fully infer the developer's intent, developers may write comments first and then trigger code completion for the corresponding code block. If the block contains multiple parallel AST nodes, AST-based masking usually covers only one node (red part in the \textbf{Figure~\ref{fig:sft-heuristic-fourth}}), leaving the remaining parts (yellow part) unmasked.

\end{itemize}

\textbf{Table~\ref{tab:ap-normal-language-statistics}} reports the language-level AP1--AP6 and Normal counts in the thematic-analysis sample. These statistics describe the in-sample distribution within the 4,000-candidate Java/Python thematic-analysis set, and should not be interpreted as population prevalence estimates for all FIM training corpora.

\begin{table}[htbp]
  \centering
  \caption{Language-Level AP and Normal Counts in the Masked Candidates for Thematic Analysis.}
  \label{tab:ap-normal-language-statistics}
  \scriptsize
  \setlength{\tabcolsep}{2.5pt}
  \renewcommand{\arraystretch}{1.2}
  \begin{tabular}{l r *{7}{>{\centering\arraybackslash}p{1.08cm}}}
    \toprule

    \textbf{Language} & \textbf{Total} & \textbf{AP1} & \textbf{AP2} & \textbf{AP3} & \textbf{AP4} & \textbf{AP5} & \textbf{AP6} & \textbf{Normal} \\

    \midrule

    Java & 2,000 & \makecell{918\\(45.9\%)} & \makecell{60\\(3.0\%)} & \makecell{5\\(0.3\%)} & \makecell{130\\(6.5\%)} & \makecell{24\\(1.2\%)} & \makecell{44\\(2.2\%)} & \makecell{819\\(41.0\%)} \\

    Python & 2,000 & \makecell{928\\(46.4\%)} & \makecell{56\\(2.8\%)} & \makecell{3\\(0.2\%)} & \makecell{46\\(2.3\%)} & \makecell{32\\(1.6\%)} & \makecell{48\\(2.4\%)} & \makecell{887\\(44.4\%)} \\

    \bottomrule
  \end{tabular}

  \vspace{0.3em}
  \parbox{0.96\linewidth}{\footnotesize \textit{Note.} AP1--AP6 and Normal are consensus results reached by the three researchers after discussion. \textbf{Normal} denotes samples that the three researchers considered not to belong to AP1--AP6 and not to exhibit an obvious anti-pattern. Percentages are calculated within the 2,000 masked candidates for each language.}
\end{table}

Through this systematic thematic analysis process, we qualitatively constructed a structured anti-pattern classification framework. We also observed typical anti-pattern examples in the ExecRepoBench~\citep{yang2024execrepobench} benchmark set, as is shown in (\textbf{Figure~\ref{fig:anti-pattern-a}} and \textbf{Figure~\ref{fig:anti-pattern-b}}).
This indicates both that we have not misinterpreted the descriptions of masking strategies in existing literature, and that the various issues (aka, anti-patterns) caused by prevailing masking strategies are a commonly reported and widespread phenomenon.

In addition to the aforementioned dataset-related anti-patterns, our early-stage training experiments~\citep{yu2025synthcoder} also reveal another critical anti-pattern:
\begin{itemize}
    \item \textbf{AP7: Erroneous context repetition.} The problem appears when a FIM model fails to produce a valid completion and instead copies prefix or suffix context at the target position. Repetition itself is not an anti-pattern: a completion may legitimately repeat nearby code when that is the correct result. Due to the strict response time requirements inherent in FIM workflows, practitioners often resort to training smaller-parameter LLMs for such tasks, and this erroneous code-echo can degrade usability while consuming extra tokens. This is consistent with the qualitative evaluation in prior FIM work, where failed infilling may copy the suffix and then lock into repeated prompt patterns when the model does not end the middle span properly~\citep{bavarian2022efficient}.
\end{itemize}

To address these anti-patterns, we adopted different strategies (as shown in \textbf{Table~\ref{tab:AP-solutions}}) to mitigate them. For \textbf{AP1}, we avoided character-delimited random masking, which is our \textbf{STG1}. For \textbf{AP2 to AP6}, we designed three FIM data construction strategies based on developer habits, i.e., \textbf{STG2}  --  Masking the rest of a single line, \textbf{STG3}  --  Masking to the end of code fragments within
parentheses, and \textbf{STG4}  --  Masking entire code fragments after comments. These data construction methods will be detailed in \textbf{Section~\ref{section:masking-strategies}}. Furthermore, for \textbf{AP7}, we employed \textbf{STG5: Context Repetition Suppression} to mitigate erroneous context repetition; details are provided in \textbf{Section~\ref{section:rejected-code-sampling}}.
Notably, the mapping between anti-patterns and strategies is not strictly one-to-one. For example, \textbf{AP2}, \textbf{AP3}, and \textbf{AP4} share the similar structural issue, i.e., AST-node masking can leave same-line trailing content outside the masked span even though developers usually expect the rest of the line to be completed as one unit. Therefore, \textbf{STG2} is used to mitigate all three cases. In the alignment stage, \textbf{STG5} is the explicit repetition-suppression strategy for \textbf{AP7}, while \textbf{STG6} is introduced as a general preference-pair construction strategy rather than as a direct remedy for a specific anti-pattern.

\begin{table}[htbp]
  \centering
  \caption{Anti-Patterns and Corresponding Solutions.}
  \label{tab:AP-solutions}
  \small
  \setlength{\tabcolsep}{4pt}
  \renewcommand{\arraystretch}{1.3}
  \begin{tabular}{c p{5cm} p{5cm}}
    \toprule

    \textbf{ID} & \textbf{Anti-pattern Type} & \textbf{Proposed Solution} \\

    \midrule

    \textbf{AP1} & Truncated identifiers and AST units & \textbf{STG1}: Avoiding character-delimited random masking \\

    \textbf{AP2} & Masked parameter AST nodes missing trailing content & \textbf{STG2}: Masking the rest of a single line \\

    \textbf{AP3} & Masked condition AST nodes missing trailing content & \textbf{STG2}: Masking the rest of a single line \\

    \textbf{AP4} & Masked expression or call AST nodes missing trailing content & \textbf{STG2}: Masking the rest of a single line \\

    \textbf{AP5} & Masked parenthesized content not preserving semantic integrity & \textbf{STG3}: Masking to the end of code fragments within parentheses  \\

    \textbf{AP6} & Masked content following comments not preserving semantic integrity & \textbf{STG4}: Masking entire code fragments after comments  \\

    \textbf{AP7} & Erroneous context repetition & \textbf{STG5}: Context repetition suppression \\

    \bottomrule
  \end{tabular}
\end{table}

\subsubsection{Optimized Masking Strategies} \label{section:masking-strategies}

To effectively address the anti-patterns and construct a better training corpus, we devised the following optimized masking strategies.

\paragraph*{\textbf{STG1: Avoiding character-delimited random masking}.} \
We deliberately avoid adopting character-delimited random masking strategies. Such approaches fail to capture the inherent syntactic structures and logical dependencies within code, often resulting in completion targets whose boundaries contain incomplete keywords or identifiers. These fragmented contexts are highly unrepresentative of real-world development practices, where developers rarely expect or benefit from such completion suggestions. We use the Tree-sitter~\footnote{\url{https://tree-sitter.github.io/tree-sitter}} parsing tool to parse code files into ASTs. Based on a variety of node extraction strategies, we select target nodes as the code snippets to be completed and extract their corresponding prefix and suffix contexts.

\paragraph*{\textbf{STG2: Masking the rest of a single line}.} \
When a developer pauses typing and there is no code to the right of the cursor, code completion is often triggered to finish the current line. We designed two strategies for this scenario.

    \begin{itemize}
        \item \textbf{Random Intra-Line Completion}: Developers may expect code completion even when a single identifier has not been fully typed. Traditional editors typically provide word-level suggestions in such cases, whereas AI-assisted coding tools more commonly complete the entire line. To simulate this behavior, we randomly select a position within a line and use the remainder of the line as the completion target (i.e., masking part).

        \item \textbf{Syntax-Aware Triggered Completion}: The completion is triggered immediately following a language-specific, syntactically significant token. We define a curated set of such tokens for each programming language, encompassing critical keywords (e.g., for control flow and definitions, see \textbf{Figure~\ref{fig:sft-heuristic-first}}) and operators (e.g., for assignment and member access, see \textbf{Figure~\ref{fig:sft-heuristic-second}}). The remainder of the line after the token is designated as the completion target.

    \end{itemize}

\paragraph*{\textbf{STG3: Masking to the end of a code fragment within parentheses}.} \
Paired parentheses often enclose relatively independent code snippets. In code editors, typing an opening parenthesis automatically generates the closing one and places the cursor in between. We sample data to replicate this common scenario, as illustrated in \textbf{Figure~\ref{fig:sft-heuristic-third}}.

\paragraph*{\textbf{STG4: Masking entire code fragments after comments}.} \
When a model cannot directly understand a developer's intent, the developer may manually write a comment describing the functionality of the code block to be written, then trigger code completion on the next line. To mimic this, we use Tree-sitter to detect full-line comment nodes in code files and treat the first code block immediately following the comment as the completion target, as illustrated in \textbf{Figure~\ref{fig:sft-heuristic-fourth}}.

\subsubsection{Rejected Code Sampling Strategies} \label{section:rejected-code-sampling}
To further optimize the model's FIM capabilities, we also introduce paired preference alignment training. The DPO-stage data construction is based on the original 500,000 samples reserved for alignment in the final training corpus, and the Seed and Qwen series use the same data split as the source for constructing preference pairs. The core of this training is to construct a high-quality preference dataset. We therefore design two complementary sources of preference pairs: targeted context-repetition suppression pairs for \textbf{STG5}, and regular sampled negative-code pairs for \textbf{STG6}.
Finally, for each retained negative sample, we build a preference pair: the sampled negative completion is the \texttt{rejected\_}\allowbreak\texttt{response}, and the corresponding canonical ground-truth middle span is the \texttt{chosen\_}\allowbreak\texttt{response}. Thus, the chosen side is always the canonical completion of the original FIM instance, while the rejected side comes from sampled model outputs retained after filtering. We anchor the chosen side to the canonical middle span because large-scale FIM training instances are often local fragments without a stable executable harness or test oracle. Therefore, STG6 relies on textual filtering rather than test-backed validation to construct scalable preference pairs.

\begin{figure}[h]
    \centering
    \begin{subfigure}[t]{0.8\textwidth}
        \centering
        \includegraphics[width=\linewidth]
        {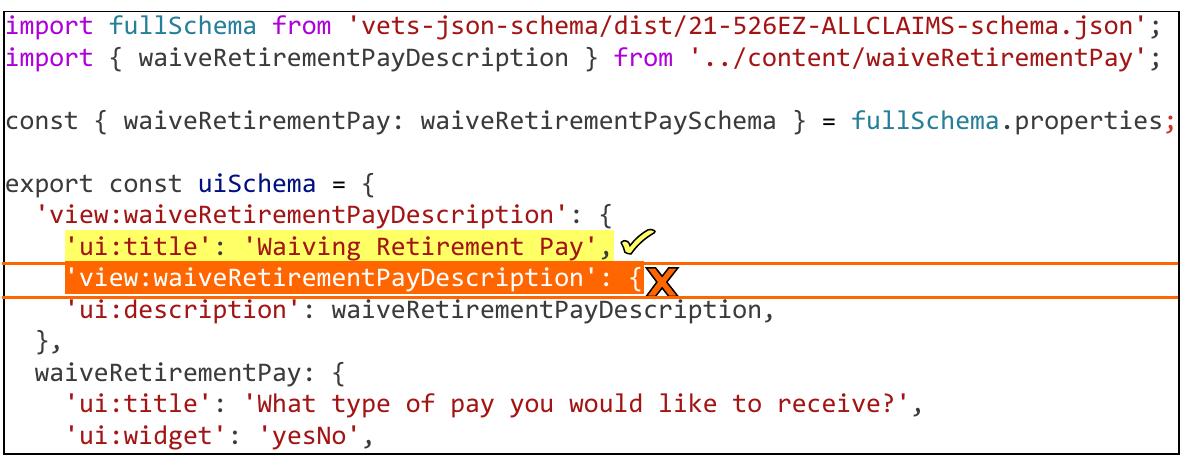}
        \caption{Example of a Completion That Exhibits Prefix Context Repetition.}
        \label{fig:prefix-rejected}
    \end{subfigure}

    \vspace{10 pt}
    \begin{subfigure}[t]{1\textwidth}
        \centering
        \includegraphics[width=0.8\linewidth]
        {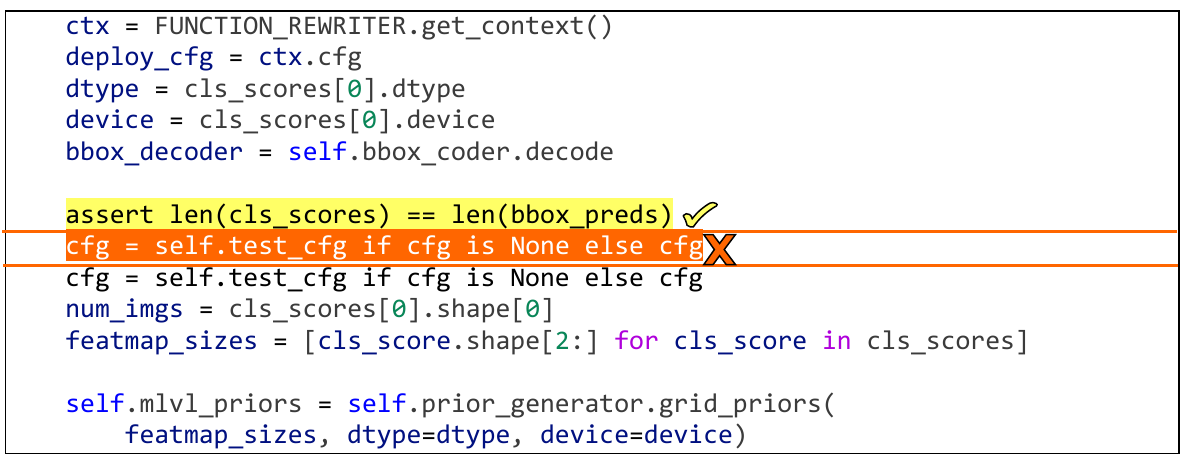}
        \caption{Example of a Completion That Exhibits Suffix Context Repetition.}
        \label{fig:suffix-rejected}
    \end{subfigure}

    \raggedright

    {
    \vspace{0.5 em}
     \textsuperscript{\dag} The \yellowBG{yellow region} denotes the ground-truth completion, while the \orangeBG{orange region} represents the observed incorrect completion, which only repeats the code  nearby.
    }

    \caption{Illustrative Examples of Prefix and Suffix Context Repetition.}
    \label{fig:rejected-2}
\end{figure}

\paragraph*{\textbf{STG5: Context repetition suppression}.} \
Models with relatively small parameter size may echo either the prefix or the suffix of the context~\citep{bavarian2022efficient}.
Here, we focus on erroneous context repetition: the model directly copies neighboring context instead of generating a meaningful and valid completion. We provide illustrative examples of prefix and suffix context repetition in \textbf{Figure~\ref{fig:prefix-rejected}} and \textbf{Figure~\ref{fig:suffix-rejected}}.

To suppress code-echo, we allocate 11\% of the DPO-stage data split to context repetition suppression. STG5 covers representative prefix/suffix neighboring-context repetition cases rather than all possible repetition forms. Specifically, suffix-context repetition accounts for 10\% of the overall DPO-stage data split. For these instances, we select cases where the expected completion after stripping does not start with the first line of the stripped suffix, and construct suppression pairs where the positive example is the ground-truth completion and the negative example is the first line of the suffix code. Prefix-context repetition accounts for 1\% of the overall DPO-stage data split and is constructed in a similar way. This textual rule excludes potential false positives where the correct completion itself is a repetition of the adjacent context line. More complex code-echo cases, such as multi-line repetition or long-span copying with small local edits, require more involved definitions and detection rules and are beyond the STG5 construction scope of this study.

\paragraph*{\textbf{STG6: General rejected code sampling for alignment training}.} \
To further support the preference alignment process, STG6 constructs regular preference pairs that help the model better distinguish canonical FIM completions from plausible but flawed completions under different masking patterns. This strategy is not designed to target a single anti-pattern; instead, it strengthens the model's general understanding of completion objectives and masking-pattern-conditioned generation. STG6 accounts for 89\% of the DPO-stage data split. Based on the DPO-stage alignment data, we use the best checkpoint from SFT to generate 10 candidate negative samples for each data point with the temperature set to 1.0.

To ensure the quality of negative samples, we implemented a series of textual safeguards on the candidate set.
First, following the method in~\citet{li2025aixcoder}, we eliminated noisy samples, such as duplicates and empty strings, and excluded samples that were identical to or contained the ground truth, so that the rejected side does not reproduce the canonical answer.
Second, since test pass rates alone cannot fully evaluate the utility of a completion, i.e., completions that are textually similar to the ground truth may still reduce the developer's editing workload~\citep{dibia2023aligning},
we employed the BLEU score as a textual similarity metric to filter out candidates that were excessively similar to the ground truth, thereby avoiding penalizing high-quality completions that are close to the canonical span.
In practice, we calculated the BLEU score for each negative sample against the corresponding ground truth and discarded samples with a score exceeding 0.9. These filtering rules are safeguards for scalable preference construction, not functional-correctness proofs that every retained rejected response is necessarily wrong.

After the filtering process, we retain up to 3 negative samples for each original data point. These sampled data are then combined with the
context repetition suppression data to form the final paired dataset for preference alignment training.

\begin{figure}[h]
    \centering
    \begin{subfigure}[t]{0.8\textwidth}
        \centering
        \includegraphics[width=\linewidth]
        {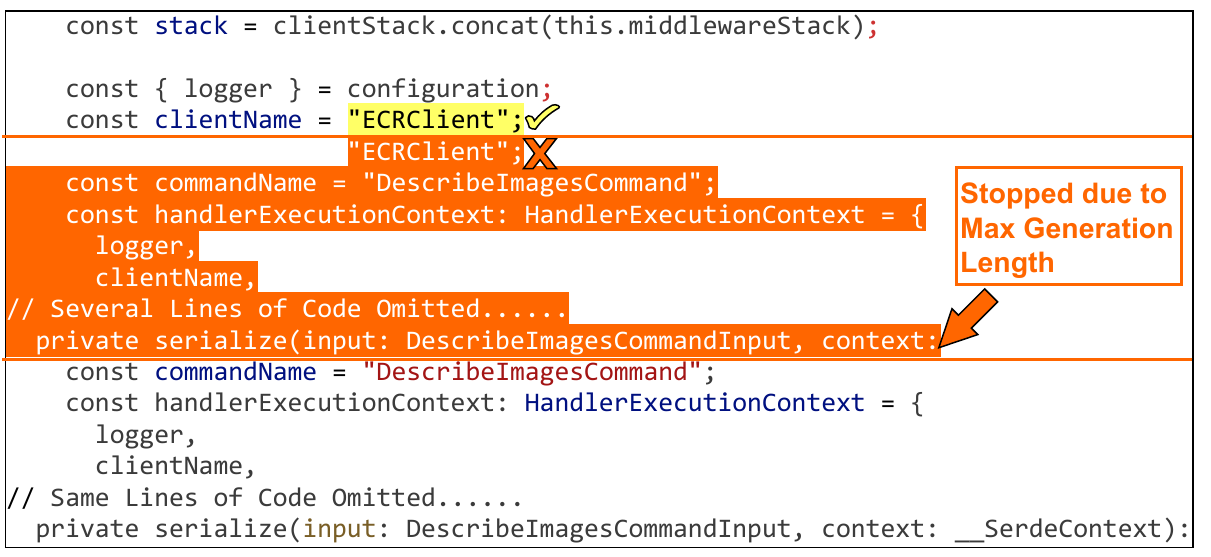}
        \caption{Example of a Pre-filter Candidate Negative Sample That Exhibits Extreme Repetition. After Correctly Producing the Ground-Truth Middle Part, the Model Repeats the Suffix Until the Token Limit Is Reached.}
        \label{fig:hit-token-limit-rejected}
    \end{subfigure}

        \vspace{10 pt}
    \begin{subfigure}[t]{0.8\textwidth}
        \centering
        \includegraphics[width=\linewidth]
        {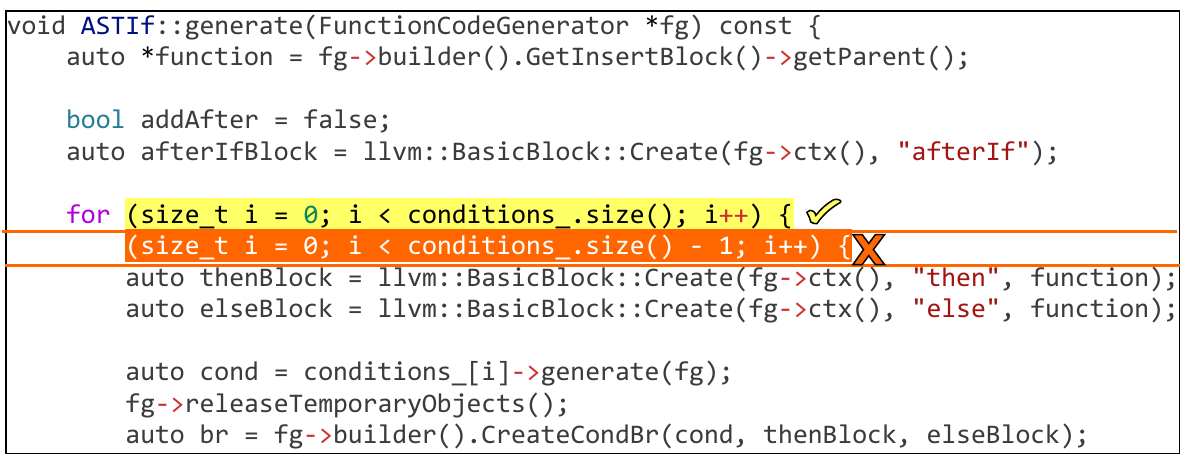}
        \caption{Example of a Conventional Negative Sample. The Model Introduces a Logical Error Without Producing Repetition.}
        \label{fig:regular-rejected}
    \end{subfigure}

    \raggedright

    {
    \vspace{0.5 em}
     \textsuperscript{\dag} The \yellowBG{yellow region} denotes the ground-truth completion; the \orangeBG{orange region} represents the sampled negative completion.
    }

    \caption{Illustrative Examples of (a) Extreme Repetition and (b) Conventional Errors in the Pre-filter Candidates.}
    \label{fig:rejected-1}
\end{figure}

During pre-filter candidate generation for STG6 negative sampling, we also observed an extreme failure mode in approximately 12.54\% of the candidate negative completions: the model repetitively generates input context until reaching the maximum output length, as shown in \textbf{Figure~\ref{fig:hit-token-limit-rejected}}. This observation motivates keeping \textbf{STG5} as a dedicated repetition-suppression component. By contrast, the regular STG6 pairs mainly consist of conventional flawed completions, such as logical or syntactic errors without extreme repetition (see \textbf{Figure~\ref{fig:regular-rejected}}), and are used to support general preference alignment for FIM completion.

\begin{figure}[!ht]
    \centering
    \includegraphics[width=1\textwidth]
    {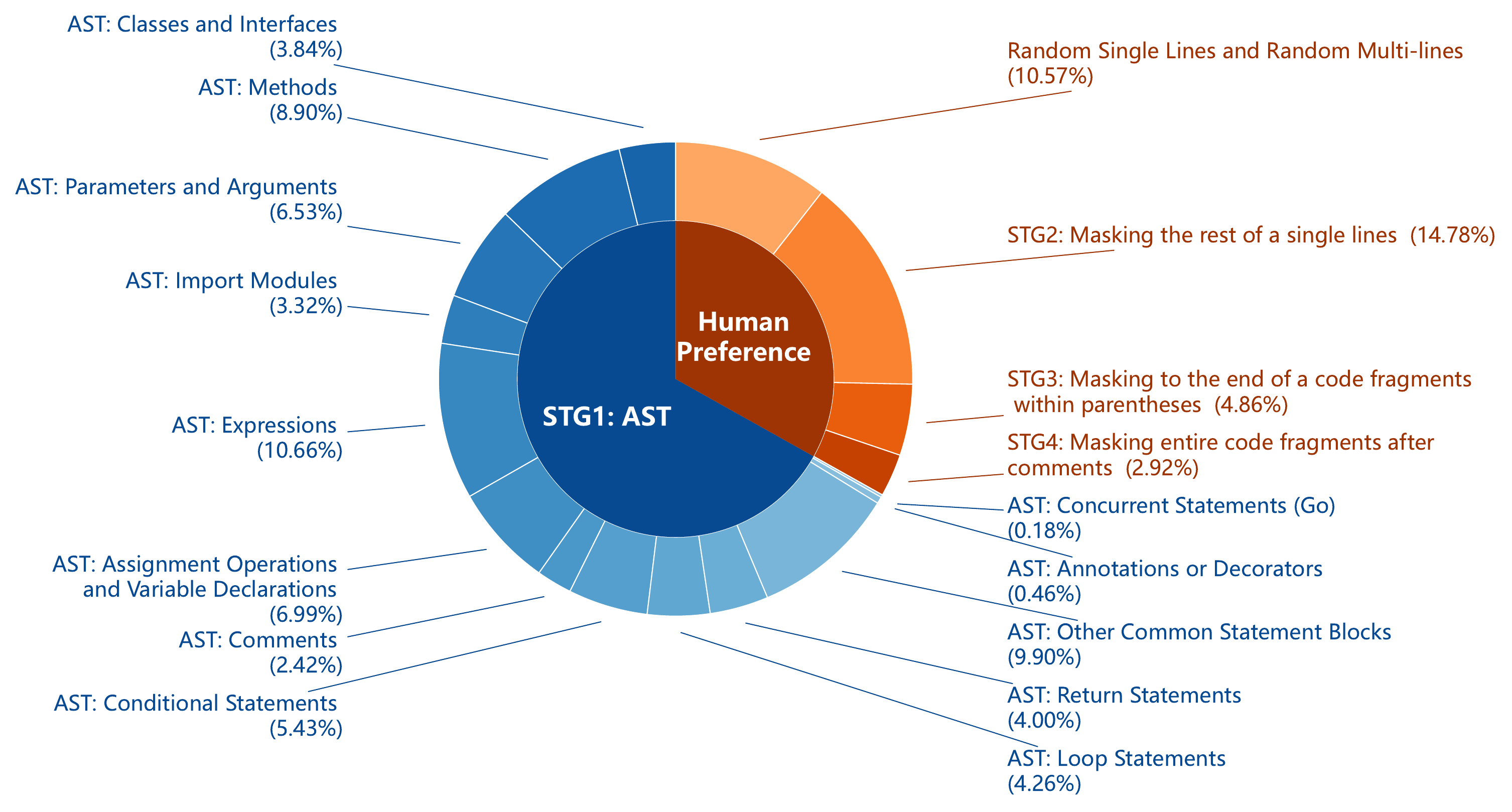}
    \caption{Composition of the Dataset by Generation Strategy.}
    \label{fig:data_composition}
\end{figure}

\subsubsection{The final training corpora}\label{section:final-training-corpora}
As a result, we constructed model-specific FIM training corpora after the PPL filtering stage. Each corpus contains 1 million FIM training samples covering six mainstream programming languages: Java, Python, JavaScript, TypeScript, C++, and Go. Following the data-allocation practice referenced from aiXcoder-V2~\citep{li2025aixcoder}, we split the constructed samples at a 1:1 ratio between fine-tuning and DPO-stage alignment, with the latter providing the original 500,000 samples used as the source data for DPO construction. Both corpora are derived from the same candidate pool and the same strategy-based optimized masking pipeline, but the final retained samples may differ because perplexity is computed with respect to different base models. Each individual strategy is backed by sufficient training data, ensuring that the finetuned model acquires robust completion capabilities across diverse coding scenarios, as illustrated in \textbf{Figure~\ref{fig:data_composition}}. It should be noted that the data distribution across these categories reflects the operational outcome of the processing procedure and filtering criteria, rather than a tuned optimum or an intentionally designed mixture.

\subsection{Model Post-training} \label{section:training_details}

We applied a two-stage post-training process based on the dataset described in the previous sections. To verify the generality of our method beyond a single model family, we employ two recent compact FIM-capable code base models from different model families: Seed-Coder-8B-Base and Qwen2.5-Coder-7B. In the SFT stage, Seed and Qwen follow the same fine-tuning method.
PPL filtering uses the base model of the corresponding model family. Thus, suffix-free Qwen variants use Qwen-specific PPL filtering, whereas the \texttt{-S} suffix marks Qwen variants trained through the Seed-PPL data pipeline. For Qwen, \CoderQwen-SFT-wo-H and \CoderQwen-SFT are the Qwen-filtered SFT variants, while \CoderQwen-SFT-S is the Seed-filtered SFT variant retained as the starting point for the Qwen DPO-stage variants, allowing them to match the Seed DPO variants in input corpora and training procedure while changing the pretrained model family from Seed to Qwen. We denote the resulting post-trained models as \CoderSeed\ and \CoderQwen\ , respectively.
During post-training, both models utilize a FIM objective; however, the specific FIM prompt templates are adapted to align with each base model's pre-training protocols. For \CoderSeed, we adopt its original SPM (Suffix-Prefix-Middle) format~\citep{seed2025seedcoder}:

\begin{quote}
\small \texttt{<[fim-suffix]>}\textit{\{SUFFIX\}}\texttt{<[fim-prefix]>}\textit{\{PREFIX\}}\texttt{<[fim-middle]>}\textit{\{MIDDLE\}}
\end{quote}

Conversely, for \CoderQwen, we utilize its orthodox PSM (Prefix-Suffix-Middle) format~\citep{hui2024qwen2}:

\begin{quote}
\small \texttt{<|fim\_prefix|>}\textit{\{PREFIX\}}\texttt{<|fim\_suffix|>}\textit{\{SUFFIX\}}\texttt{<|fim\_middle|>}\textit{\{MIDDLE\}}
\end{quote}

In both cases, the placeholders \textit{\{SUFFIX\}}, \textit{\{PREFIX\}}, and \textit{\{MIDDLE\}} represent the corresponding code segments
, while the rest tags denote model-specific special tokens.
Besides, with reference to~\cite{gong2025structure}, we set FIM rate to 0.7.

\subsubsection{Fine-tuning Stage} \label{section:finetune_stage}

In the fine-tuning stage, we adopted the curriculum learning method for FIM tasks from~\cite{sagtani2025improving}.

\paragraph*{\uline{Curriculum Learning.}} \
The data preprocessing for curriculum learning fine-tuning is primarily divided into two steps:
\begin{itemize}
  \item \textbf{Step 1: Complexity Sorting.} \\
        We use Tree-sitter to parse code files and generate ASTs, then measure complexity by the number of identifier nodes in the AST and sort them in descending order as is described in~\cite{sagtani2025improving}. Symbols are extracted by recursively traversing the ASTs.

  \item \textbf{Step 2: Sample Selection.} \\
        We select the top $k\%$ (k=30) most complex fragments to serve as FIM and completion training samples, respectively, forming the final training dataset.
\end{itemize}

\paragraph*{\uline{Training Details.}} \
In the curriculum learning fine-tuning stage, we performed full-parameter fine-tuning on the Seed-Coder-8B-Base and Qwen2.5-Coder-7B models, updating all of their weights. This stage serves to establish the model's basic FIM task competence before the subsequent preference alignment stage. The model was trained for one epoch using the AdamW optimizer with a learning rate of $3 \times 10^{-7}$ and a 50-step linear warmup. The training data consisted of the top 30\% ($k=0.3$) most complex samples from the original dataset, with all input sequences truncated to a maximum length of 8192 tokens. By prioritizing structurally more complex samples, the fine-tuning stage emphasizes completion scenarios that require stronger structural understanding and longer-range contextual dependency modeling. The training was conducted on four NVIDIA H800 (80GB PCIe) GPUs, using a global batch size of 256 and a micro-batch size of 1 per GPU. To enhance efficiency, we employed several optimization strategies, including DeepSpeed ZeRO-1~\citep{ZeRO}, Flash Attention~\citep{dao2022flashattention}, and gradient checkpointing. Under this setup, checkpoints are recorded every 100 steps, corresponding to approximately 2.5 hours of fine-tuning per checkpoint interval; only the best checkpoint is retained for the subsequent alignment stage.

\subsubsection{Alignment Stage}\label{section:align-stage}

To make the code generation model more accurately align with the intentions and preferences of human developers, we use the DPO algorithm~\citep{rafailov2023direct} to align the model during the preference alignment stage,
based on the pairs of \texttt{chosen\_response}
and \texttt{rejected\_response} constructed in \textbf{Section~\ref{section:rejected-code-sampling}}.

\paragraph*{\uline{DPO Algorithm.}} \
Compared to the traditional RLHF framework, which relies on an explicit reward model and reinforcement learning, DPO provides a more stable and efficient end-to-end optimization path. The core idea of DPO is to directly transform the optimization objective on a preference dataset into a maximum likelihood estimation problem for the language model.

Given a preference dataset $\mathcal{D} = \{(x^i, y_w^i, y_l^i)\}$ consisting of a prompt $x$, a winning code completion $y_w$, and a losing code completion $y_l$, the goal of DPO is to finetune the policy model $\pi_{\theta}$ to better rank human preferences. The DPO algorithm optimizes the model by minimizing the following loss function:

\begin{equation*}
\begin{aligned}
\mathcal{L}(\pi_{\theta}; \pi_{\text{ref}}) = -\mathbb{E}_{(x, y_w, y_l) \sim \mathcal{D}} \Bigg[\log \sigma \left( \beta \log \frac{\pi_{\theta}(y_w|x)}{\pi_{\text{ref}}(y_w|x)} - \beta \log \frac{\pi_{\theta}(y_l|x)}{\pi_{\text{ref}}(y_l|x)} \right) \Bigg]
\end{aligned}
\end{equation*}

\begin{figure*}[h]
    \centering
    \begin{subfigure}[t]{0.32\textwidth}
        \centering
        \includegraphics[width=\textwidth]
        {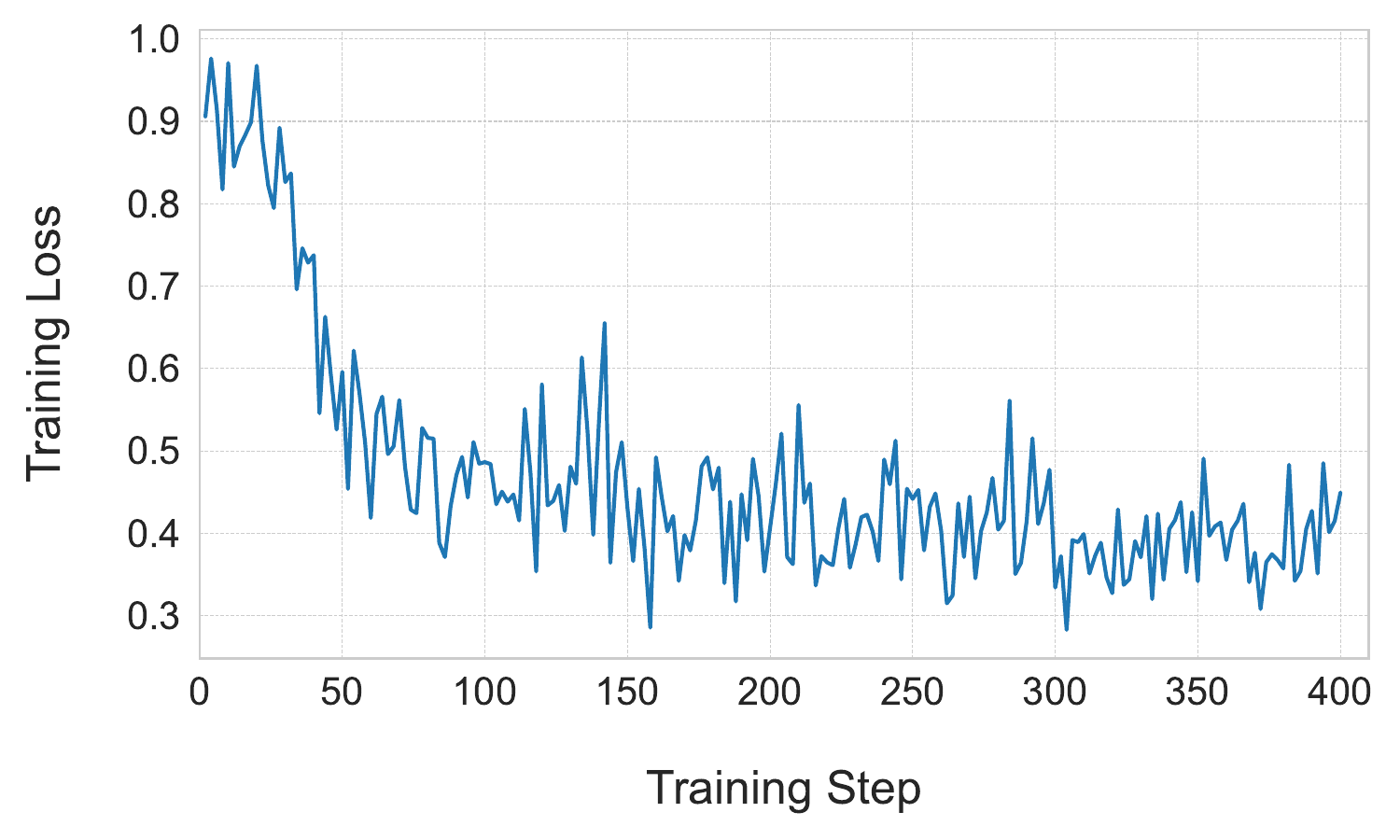}
        \caption{Training Loss Curve of \CoderSeed\ .}
        \label{fig:a}
    \end{subfigure}\hfill
    \begin{subfigure}[t]{0.32\textwidth}
        \centering
        \includegraphics[width=\textwidth]
        {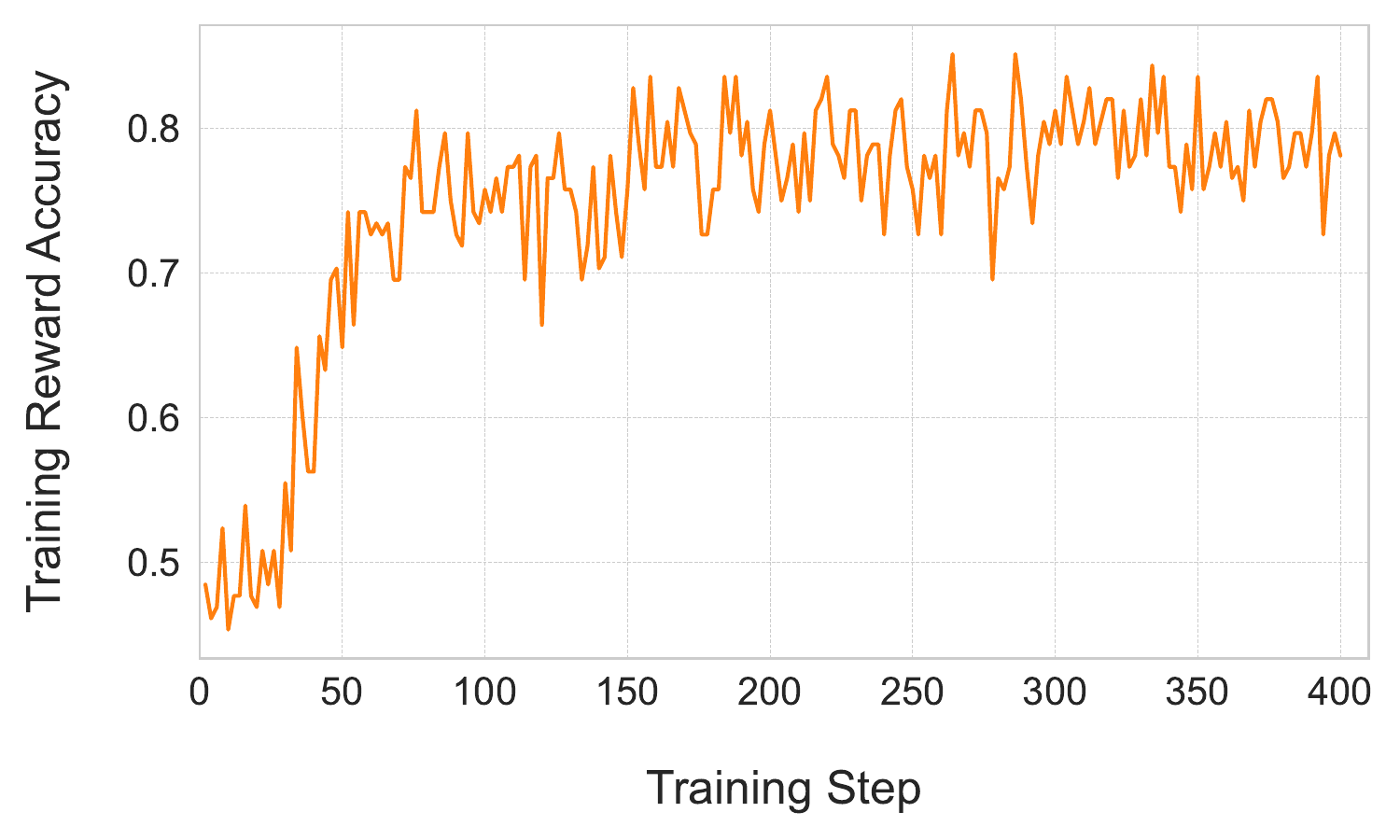}
        \caption{Training Reward Accuracy Curve of \CoderSeed\ .}
        \label{fig:b}
    \end{subfigure}\hfill
    \begin{subfigure}[t]{0.32\textwidth}
        \centering
        \includegraphics[width=\textwidth]
        {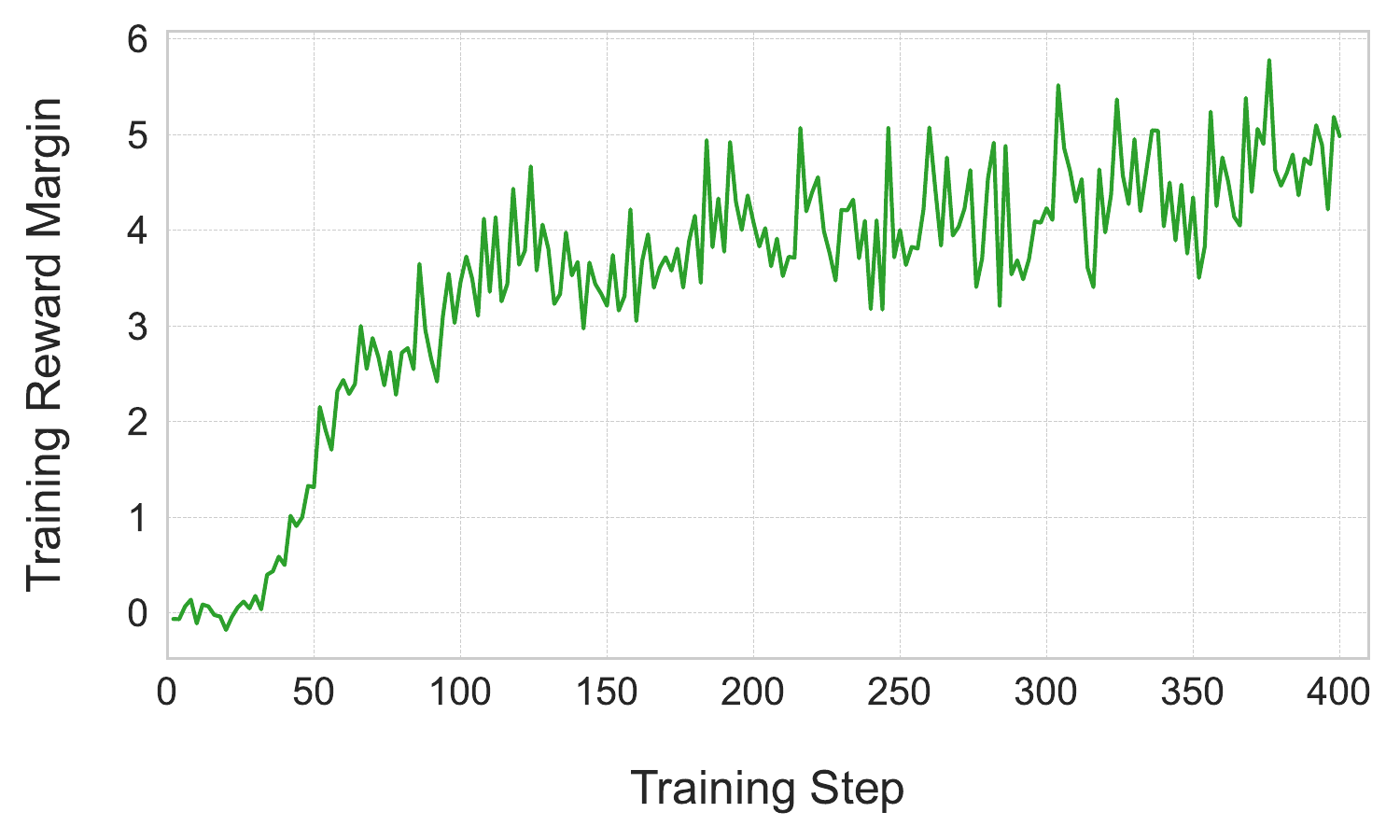}
        \caption{Training Reward Margin Curve of \CoderSeed\ .}
        \label{fig:c}
    \end{subfigure}
    \begin{subfigure}[t]{0.32\textwidth}
        \centering
        \includegraphics[width=\textwidth]
        {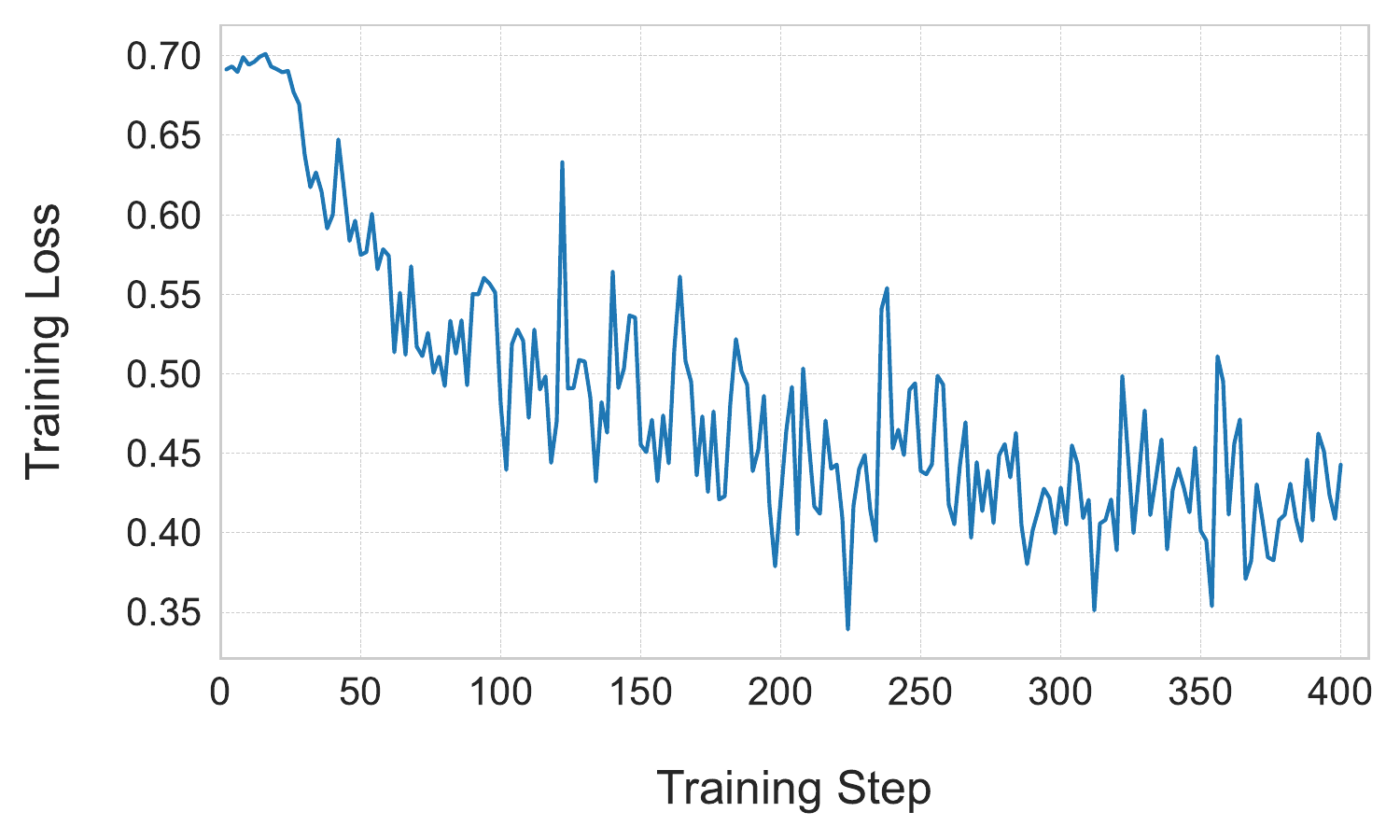}
        \caption{Training Loss Curve of \CoderQwen\ .}
        \label{fig:d}
    \end{subfigure}\hfill
    \begin{subfigure}[t]{0.32\textwidth}
        \centering
        \includegraphics[width=\textwidth]
        {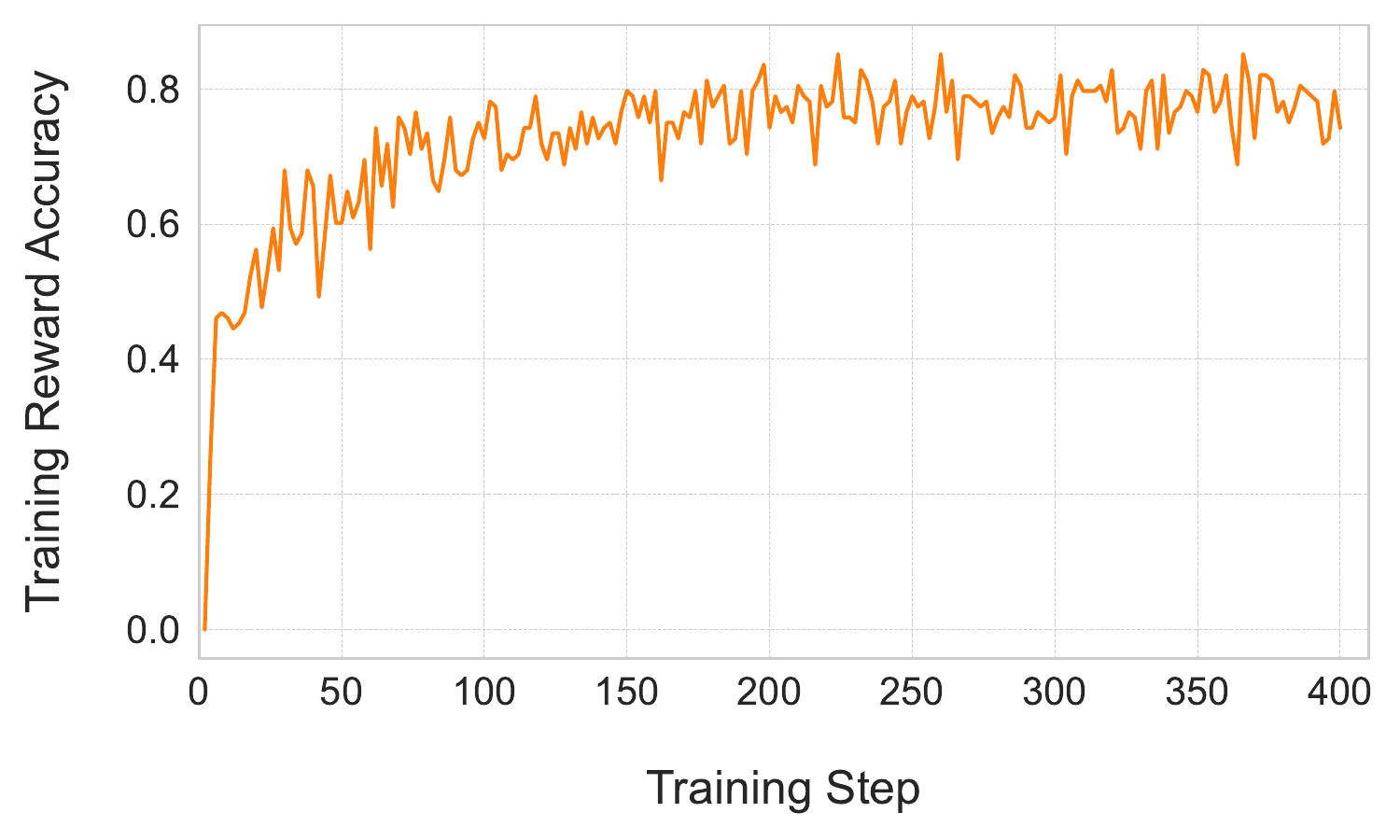}
        \caption{Training Reward Accuracy Curve of \CoderQwen\ .}
        \label{fig:e}
    \end{subfigure}\hfill
    \begin{subfigure}[t]{0.32\textwidth}
        \centering
        \includegraphics[width=\textwidth]
        {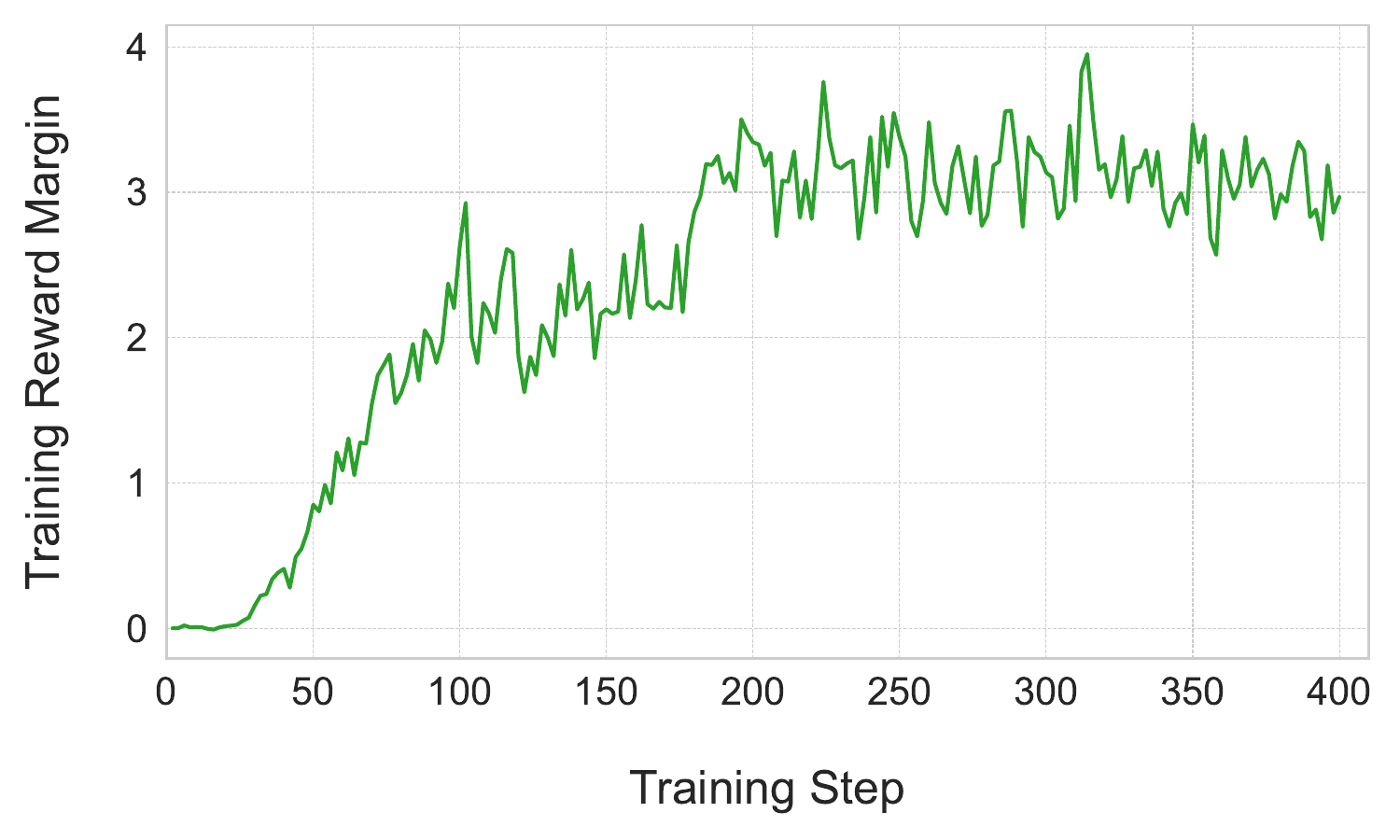}
        \caption{Training Reward Margin Curve of \CoderQwen\ .}
        \label{fig:f}
    \end{subfigure}
    \caption{Curves in the DPO Training Process of \Coder.}
    \label{fig:dpo-figures}
\end{figure*}

\paragraph*{\uline{Training Details.}} \
In the preference alignment stage using the DPO algorithm, the DPO parameter $\beta$ was set to 0.9, with an equivalent batch size of 64. The model was trained for one epoch with a maximum learning rate of 1e-4, a linear learning rate warmup of 32 steps, and a cosine learning rate decay scheduler. The maximum length for each training data was 8192 tokens.
For parameter-efficient fine-tuning, we employed the Low-Rank Adaptation (LoRA) technique~\citep{hu2022lora}. The LoRA $rank$ was set to 32 and the scaling factor $alpha$ was set to 64.
The training was conducted on 4 NVIDIA H800 80GB PCIe GPUs, leveraging the TRL framework~\footnote{\url{https://github.com/huggingface/trl}} (version 0.17.0) in conjunction with DeepSpeed ZeRO-2~\citep{ZeRO} (with a CPU offloading strategy) and gradient checkpointing.
\textbf{Figure~\ref{fig:dpo-figures}} illustrates the training trajectories of Loss, Reward Accuracy, and Reward Margin during the process of the finetuned model from Seed-Coder-8B-Base and Qwen2.5-Coder-7B.
Under this setup, checkpoints are recorded every 100 steps, corresponding to roughly 6 hours of alignment per checkpoint interval; only the best checkpoint will be retained. It can be observed that the Training Loss decreased relatively steadily, while the Reward Accuracy and Reward Margin increased and then remained relatively stable.

\section{Evaluation}\label{section:evaluation}

To evaluate \Coder\, we conducted a comprehensive evaluation on multiple code completion benchmarks against a curated set of baseline models as well as several variants for the ablation study. Our evaluation extends the \textit{qwencoder-eval}~\footnote{\url{https://github.com/QwenLM/Qwen3-Coder/tree/main/qwencoder-eval}} framework, which guarantees that our evaluation process is aligned with established community standards.

\begin{figure}[htb]
    \centering
    \includegraphics[width=0.99\textwidth]{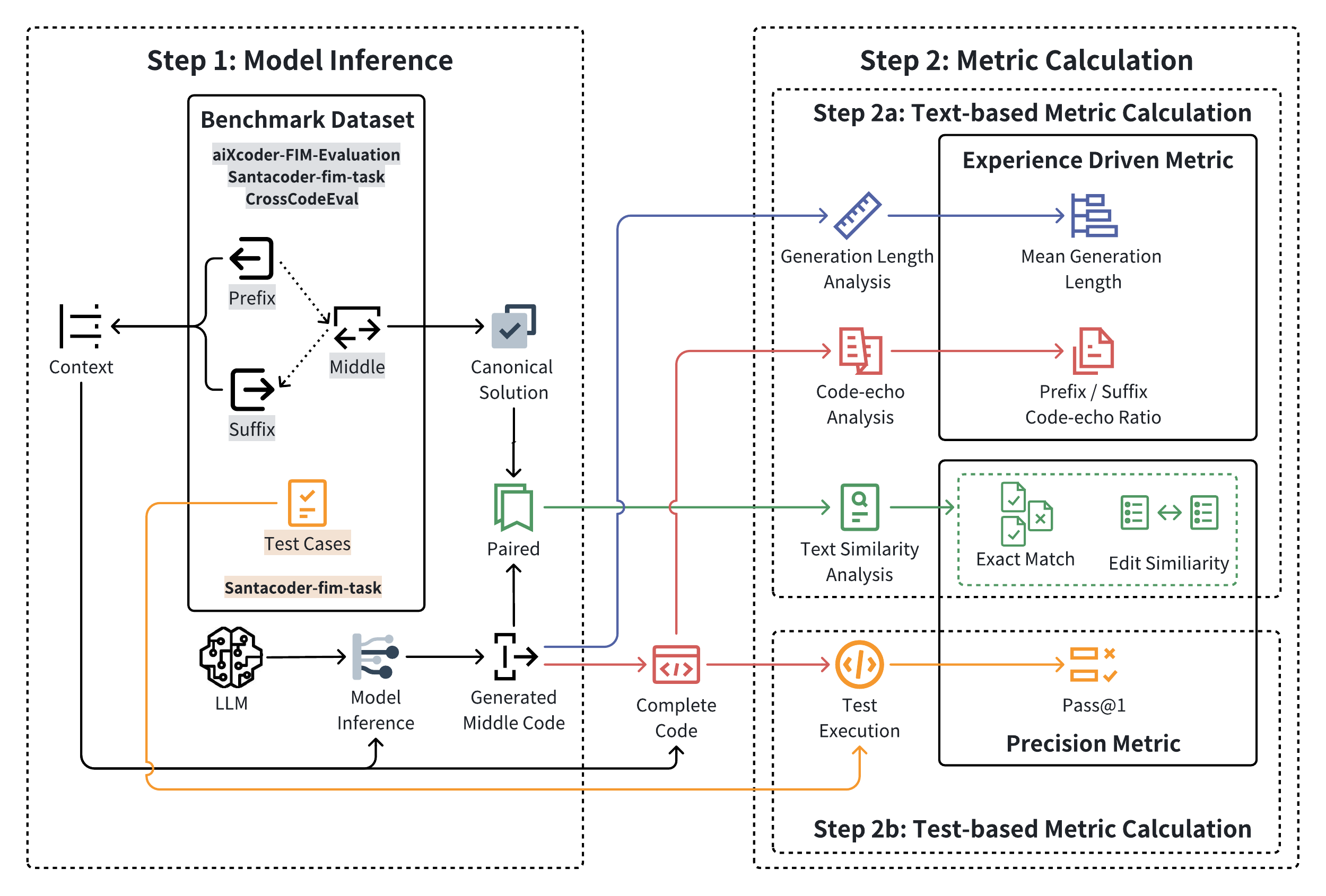}
    \caption{The Evaluation Process of \Coder.}
    \label{fig:evaluation_process}
\end{figure}

As illustrated in \textbf{Figure~\ref{fig:evaluation_process}}, the evaluation pipeline for \Coder\ comprises two primary phases: \textbf{Step 1: Model Inference} and \textbf{Step 2: Metric Calculation}. During \textbf{Step 1}, the inference phase, each model generates the \textit{\{MIDDLE\}} code segment using model-specific FIM prompt formats (detailed in \textbf{Section~\ref{section:baselines}}). All inferences are executed on a single H800 80GB PCIe GPU utilizing vLLM~\footnote{\url{https://github.com/vllm-project/vllm}} v0.9.1. Following generation, the evaluation diverges into two parallel substeps to assess performance; the detailed definitions for the specific metrics used in this phase are provided in \textbf{Section~\ref{section:metrics}}. \textbf{Step 2a} computes text-based metrics: it assesses Precision Metrics (EM, ES) by comparing the generated code to the canonical solution, and Experience Driven Metrics (Mean Generation Length, Code-echo Ratio) by analyzing the generated output relative to the input context. For benchmarks with executable test cases, \textbf{Step 2b} computes test-based metrics; here, the generated middle code is merged with the original prefix and suffix to reconstruct the complete code, which is then executed against test cases to determine the Pass@1 score.

\subsection{Benchmarks}

We select Santacoder-fim-task~\footnote{\url{https://huggingface.co/datasets/bigcode/Santacoder-fim-task}} to evaluate functional correctness with Pass@1, and aiXcoder-FIM-Evaluation~\footnote{\url{https://github.com/aixcoder-plugin/aiXcoder-FIM-Evaluation}} and CrossCodeEval~\footnote{\url{https://github.com/amazon-science/cceval}} to evaluate similarity with canonical answers.
These benchmarks are used only for evaluation; their repository or snippet provenance is also used for the leakage-control procedure described in \textbf{Section~\ref{section:data-sources}}.

\subsubsection{Santacoder-fim-task}

To assess functional correctness, we adopt Santacoder-fim-task~\citep{allal2023santacoderdontreachstars}, an extension of OpenAI's human-eval-infilling~\footnote{\url{https://github.com/openai/human-eval-infilling}} with test cases for Java ($n=1,780$), Python ($n=1,043$), and JavaScript ($n=1,969$), which is one of the most popular benchmark dataset used in multiple studies~\citep{sagtani2025improving, jiang2024aixcoder}. Its focus on short, single-line completions also makes it effective for evaluating a model's ability to avoid over-generation.

For each data instance, the model-generated code infill is concatenated with its corresponding prefix and suffix to construct a complete program. This program is then executed against the provided test cases to determine its validity. The primary evaluation metric is the Pass@1, calculated individually for each of the language. All tests were conducted on a server running Ubuntu 20.04 LTS, equipped with dual Intel(R) Xeon(R) Platinum 8468V CPUs and 1.5 TB of memory. The software stack included OpenJDK 11, Python 3.10, and Node.js 18.20.5. During the evaluation, tasks were executed with a concurrency of 32 and a timeout of 10 seconds per task.

\subsubsection{aiXcoder-FIM-Evaluation}

To evaluate similarity between target completion and model generation, we utilized aiXcoder-FIM-Evaluation~\citep{jiang2024aixcoder}, which contains 8,000 samples each for C++, JavaScript, Java, and Python, totaling 32,000 samples. It evaluates FIM code completion on files from real-world open-source projects in a style that is claimed to be closer to actual completion scenarios.

\subsubsection{CrossCodeEval}
To include an unseen-language view in evaluation, we use CrossCodeEval~\citep{ding2023crosscodeeval}, which is a benchmark designed to assess cross-file code completion. It uses static analysis to construct cross-file completion instances from projects in Python ($n=2,665$), Java ($n=2,139$), TypeScript ($n=3,356$), and C\# ($n=1,768$), totaling 9,928 instances that are only solvable with external context.
Notably, C\# is absent from the post-training corpora used for \Coder.
For our evaluation, we use the original context retrieved via the BM25 algorithm. This context includes the top 5 most relevant code snippets, each truncated to a maximum of 10 lines.

\subsection{Baselines}\label{section:baselines}

Our model selection focused on open-source models to ensure a fair comparison and accommodate regulation restrictions and resource constraints.
Simultaneously, to systematically evaluate the contribution of each training stage and dataset components including AST node masking (\textbf{STG1}), human behavior pattern simulation (\textbf{STG2, STG3 and STG4, abbreviated as STG2-STG4}), rejected code sampling (\textbf{STG6}), and repetition suppression (\textbf{STG5}), we developed several model variants,
starting from Seed-Coder-8B-Base and Qwen2.5-Coder-7B,
as is elaborated in \textbf{Section~\ref{section:variants}} for ablation studies.
Through comparison with baseline models and analysis of ablation variants, we comprehensively validated the contribution of each training strategy to both the precision and user experience of code completion.

\subsubsection{Open-source models}

Candidate baseline open-source models are filtered based on the following criteria:
\begin{itemize}
    \item \textbf{Model Size:} Active parameters between 6B and 8B, comparable to the \Coder\ series.
    \item \textbf{Capability:} Native support for FIM tasks with explicit prompt format in code repositories or documents.
    \item \textbf{Recency:} Released between 2024 and the first half of 2025.
\end{itemize}

Consequently, we excluded older models like InCoder-6B~\footnote{\url{https://huggingface.co/facebook/incoder-6B}} (released in 2022), CodeLlama-7b-hf~\footnote{\url{https://huggingface.co/codellama/CodeLlama-7b-hf}} (released in 2023), and models without clear native FIM support. We also evaluated instruction-tuned counterparts during the initial screening where available. However, base models are generally more suitable for FIM tasks, and our experiments show that instruction-tuned models perform poorly on most precision and generation-behavior metrics under orthodox FIM prompting. Therefore, the main comparisons below focus on native FIM-capable base models and our post-trained variants.
Finally, the set of baseline models are listed below.

\begin{itemize}

    \item \textbf{aiXcoder Series} \\
        {aiXcoder-7b~\footnote{\url{https://huggingface.co/aiXcoder/aixcoder-7b}}} is a 7B-parameter code LLM designed for real-world code generation contexts.
        {aiXcoder-7b-v2~\footnote{\url{https://huggingface.co/aiXcoder/aixcoder-7b-v2}}} is an updated 7B-parameter model optimized for long-context information and cross-file code completion.

    \item \textbf{CodeGemma Series} \\
        {CodeGemma-7B~\footnote{\url{https://huggingface.co/google/codegemma-7b}}} is a 7B-parameter base code LLM developed by Google.

    \item \textbf{Granite Series} \\
        {Granite-3.3-8B-Base~\footnote{\url{https://huggingface.co/ibm-granite/granite-3.3-8b-base}}} is an 8B-parameter base code LLM developed by IBM.

    \item \textbf{StarCoder2 Series} \\
        {StarCoder2-7B~\footnote{\url{https://huggingface.co/bigcode/starcoder2-7b}}} is a 7B-parameter code LLM from the BigCode project, trained on the The-Stack-V2 dataset.

    \item \textbf{DeepSeek-Coder Series} \\
        {DeepSeek-Coder-6.7B-Base~\footnote{\url{https://huggingface.co/deepseek-ai/deepseek-coder-6.7b-base}}} is a 6.7B-parameter base code LLM.
        {DeepSeek-Coder-V2-Lite-Base~\footnote{\url{https://huggingface.co/deepseek-ai/DeepSeek-Coder-V2-Lite-Base}}} is a lightweight Mixture-of-Experts (MoE) base code LLM (16B total, 2.4B active parameters).

   \item \textbf{Qwen-Coder Series} \\
        {Qwen2.5-Coder-7B~\footnote{\url{https://huggingface.co/Qwen/Qwen2.5-Coder-7B}}} is a 7B-parameter base code LLM with native FIM support.

    \item \textbf{Seed-Coder Series} \\
        {Seed-Coder-8B-Base~\footnote{\url{https://huggingface.co/ByteDance-Seed/Seed-Coder-8B-Base}}} is an 8B-parameter base LLM developed using model-driven data selection.

\end{itemize}

\subsubsection{\Coder\ Variants for Ablation Study }\label{section:variants}

To systematically evaluate the contribution of each training stage and dataset components, we developed several model variants,
detailed in \textbf{Table~\ref{tab:model_comparison}}.

\begin{table}[H]
  \centering
  \begin{threeparttable}
    \scriptsize
    \caption{Comparisons Between \Coder\ and Its Variants for Ablation Study.}
    \label{tab:model_comparison}
    \setlength{\tabcolsep}{0pt}
    \begin{tabular}{l>{\centering\arraybackslash}p{0.95cm}*{4}{>{\centering\arraybackslash}p{1.82cm}}}
      \toprule

      \multirow{5}{*}{\textbf{Model}} & \multirow{5}{*}{\makecell{\textbf{PPL} \\ \textbf{Model}}} & \multicolumn{2}{c}{\makecell{\textbf{Fine-tuning Dataset}
      }} & \multicolumn{2}{c}{\makecell{\textbf{Alignment Dataset}
      }} \\

      \cmidrule(lr){3-4} \cmidrule(lr){5-6}

      & & \parbox[c]{1.82cm}{\centering\textbf{AST}\\\textbf{Node}\\\textbf{(STG1)}} & \parbox[c]{1.82cm}{\centering\textbf{Human}\\\textbf{Expect.}\\\textbf{(STG2-STG4)}} & \parbox[c]{1.82cm}{\centering\textbf{Rejected}\\\textbf{Code}\\\textbf{Samp.}\\\textbf{(STG6)}} & \parbox[c]{1.82cm}{\centering\textbf{Repetition}\\\textbf{Supp.}\\\textbf{(STG5)}} \\

      \midrule

      Seed-Coder-8B-Base      & \textbf{\DR{N/A}} & \textbf{\DR{N/A}}  & \textbf{\DR{N/A}}  & \textbf{\DR{N/A}}  & \textbf{\DR{N/A}}  \\

      \SCSeed-SFT-wo-H      & Seed & \textbf{\DG{YES}} & \textbf{\DO{NO}}  & \textbf{\DR{N/A}}  & \textbf{\DR{N/A}}  \\

      \SCSeed-SFT           & Seed & \textbf{\DG{YES}} & \textbf{\DG{YES}} & \textbf{\DR{N/A}}  & \textbf{\DR{N/A}}  \\

      \SCSeed-DPO-wo-RS\tnote{*} & Seed & \textbf{\DG{YES}} & \textbf{\DG{YES}} & \textbf{\DG{YES}} & \textbf{\DO{NO}}  \\

      \textbf{\SCSeed-DPO}\tnote{*}    & Seed & \textbf{\DG{YES}} & \textbf{\DG{YES}} & \textbf{\DG{YES}} & \textbf{\DG{YES}} \\

      \midrule

      Qwen2.5-Coder-7B      & \textbf{\DR{N/A}} & \textbf{\DR{N/A}}  & \textbf{\DR{N/A}}  & \textbf{\DR{N/A}}  & \textbf{\DR{N/A}}  \\

      \SCQwen-SFT-wo-H      & Qwen & \textbf{\DG{YES}} & \textbf{\DO{NO}}  & \textbf{\DR{N/A}}  & \textbf{\DR{N/A}}  \\

      \SCQwen-SFT           & Qwen & \textbf{\DG{YES}} & \textbf{\DG{YES}} & \textbf{\DR{N/A}}  & \textbf{\DR{N/A}}  \\

      \SCQwen-SFT-S           & Seed & \textbf{\DG{YES}} & \textbf{\DG{YES}} & \textbf{\DR{N/A}}  & \textbf{\DR{N/A}}  \\

      \SCQwen-DPO-wo-RS-S\tnote{$\dagger$} & Seed & \textbf{\DG{YES}} & \textbf{\DG{YES}} & \textbf{\DG{YES}} & \textbf{\DO{NO}}  \\

      \textbf{\SCQwen-DPO-S}\tnote{$\dagger$}    & Seed & \textbf{\DG{YES}} & \textbf{\DG{YES}} & \textbf{\DG{YES}} & \textbf{\DG{YES}} \\

      \bottomrule
    \end{tabular}

    \begin{tablenotes}
      \item[] \SC\ abbreviates \Coder\ in table model names.
      \item[] Expect., Samp., and Supp. abbreviate Expectation, Sampling, and Suppression, respectively.
      \item[*] These models are based on \SCSeed-SFT.
      \item[$\dagger$] These Qwen DPO rows are initialized from \SCQwen-SFT-S and use the Seed-PPL data pipeline, matching the Seed DPO rows in input corpora and training procedure while changing the pretrained model family from Seed to Qwen.
      \item[] \textbf{\DR{N/A}}: Indicates the model did not undergo the corresponding training stage.
      \item []  \textbf{\DO{NO}}: Means that the specific type of data is not included in this training stage.
    \end{tablenotes}
  \end{threeparttable}
\end{table}

In general, our ablation study consideration is twofold.
First, in the fine-tuning stage, Seed and Qwen replicate the same SFT method. We compare \CoderSeed-SFT-wo-H and \CoderQwen-SFT-wo-H against \CoderSeed-SFT and \CoderQwen-SFT, respectively, to measure the impact of the dataset derived from \textbf{STG2-STG4} based on human behavior patterns, as described in \textbf{Section~\ref{section:masking-strategies}}. Qwen rows without the \texttt{-S} suffix use Qwen-specific PPL filtering, so this SFT-stage comparison evaluates the heuristic masking rules under a model-matched filtering setup.
Second, in the alignment stage, we intentionally keep the Qwen DPO variants on the Seed-PPL data pipeline and initialize them from \CoderQwen-SFT-S. This design pairs \CoderQwen-DPO-wo-RS-S and \CoderQwen-DPO-S with the corresponding Seed DPO variants under the same SFT/DPO input corpora and training procedure, while changing the pretrained model family from Seed to Qwen. Within this cross-model-family matched-data setting, the comparison from \CoderQwen-DPO-wo-RS-S to \CoderQwen-DPO-S is used to assess the effect of the context repetition suppression data in \textbf{STG5} (detailed in \textbf{Section~\ref{section:rejected-code-sampling}}), rather than to claim a full DPO rerun on the Qwen-filtered SFT corpus.

\subsubsection{FIM Prompt Templates}

For in-file benchmarks, such as {Santacoder-fim-task} and {aiXcoder-FIM-Evaluation}, templates are categorized by their context ordering -- PSM (Prefix-Suffix-Middle) or SPM (Suffix-Prefix-Middle) -- as detailed in \textbf{Table~\ref{tab:fim_templates}}~\citep{bavarian2022efficient}.
To the best of our knowledge, our implementation strictly adheres to the prompt formats prescribed by the respective model authors. In practice, the predicted \textit{\{MIDDLE\}} fragment is generated immediately following the FIM templates without any additional tokens. Notably, the $\hookrightarrow$ symbols in \textbf{Table~\ref{tab:fim_templates}} are purely for typographical clarity within the table rather than any actual tokens in practice.

\begin{table}[htbp]
  \centering
  \begin{threeparttable}
  \caption{FIM Prompt Template of Baseline Models.}
  \label{tab:fim_templates}
  \small
  \setlength{\tabcolsep}{2pt}
  \begin{tabular}{c >{\raggedright\arraybackslash}p{0.6\textwidth} >{\centering\arraybackslash}p{0.25\textwidth}}
    \toprule

    \textbf{Type} & \multicolumn{1}{c}{\textbf{FIM Template Format}} & \textbf{Models} \\

    \midrule

    \multirow{8}{*}{{PSM}}
    & \texttt{<|fim\_prefix|>\textit{\{PREFIX\}}<|fim\_suffix|>\textit{\{SUFFIX\}}\newline
    $\hookrightarrow$<|fim\_middle|>} &
    \makecell[t]{
    CodeGemma Series, \\ Qwen-Coder Series, \\ \SCQwen\ Variants } \\

    \cmidrule{2-3}

    & \texttt{<fim\_prefix>\textit{\{PREFIX\}}<fim\_suffix>\textit{\{SUFFIX\}}\newline
    $\hookrightarrow$<fim\_middle>} &
    \makecell[t]{
    Granite Series, \\ StarCoder2 Series
    } \\

    \cmidrule{2-3}

    & \texttt{
    <{\promptbar}fim{\promptscore}begin{\promptbar}>\textit{\{PREFIX\}}<{\promptbar}fim{\promptscore}hole{\promptbar}>\textit{\{SUFFIX\}}\newline
    $\hookrightarrow$<{\promptbar}fim{\promptscore}end{\promptbar}>}
    &
    \makecell[t]{
    DeepSeek-Coder Series
    } \\

    \midrule

    \multirow{4}{*}{{SPM}}
    & \texttt{<s>\promptscore<AIX-SPAN-PRE>{\promptscore}<AIX-SPAN-POST>\textit{\{SUFFIX\}}\newline
    $\hookrightarrow$\promptscore<AIX-SPAN-MIDDLE>\textit{\{PREFIX\}}} &
    \makecell[t]{
    aiXcoder Series\tnote{*}
    } \\

    \cmidrule{2-3}

    & \texttt{<[fim-suffix]>\textit{\{SUFFIX\}}<[fim-prefix]>\textit{\{PREFIX\}}\newline
    $\hookrightarrow$<[fim-middle]>} &
    \makecell[t]{
    Seed-Coder Series, \\ \SCSeed\ Variants } \\

    \bottomrule
  \end{tabular}
  \begin{tablenotes}
    \item[*]  The $\hookrightarrow$ symbol is purely for typographical clarity within the table rather than actually a new line in practice.
  \end{tablenotes}
  \end{threeparttable}
\end{table}

For cross-file benchmark such as CrossCodeEval, we adopt the following prompt organization strategies, strictly adhering to the original definitions provided by the authors of the models, specifically, the following strategies are adopted: the \textbf{aiXcoder Series}~\footnote{\url{https://github.com/aixcoder-plugin/aixcoder-7b-v2/blob/main/prompt/prompt\_aixcoder\_colt.py}}, \textbf{Qwen-Coder Series}~\footnote{\url{https://github.com/QwenLM/Qwen3-Coder/blob/98dee50c0618687f55e85dde0d9a652592d25053/README.md}}, \textbf{StarCoder2 Series}~\footnote{\url{https://arxiv.org/pdf/2402.19173}}, \textbf{DeepSeek Series}~\footnote{\url{https://github.com/deepseek-ai/DeepSeek-Coder/blob/main/README.md}}, \textbf{CodeGemma Series~\footnote{\url{https://ai.google.dev/gemma/docs/codegemma/prompt-structure}}} and \textbf{Seed-Coder Series}~\footnote{\url{https://github.com/ByteDance-Seed/Seed-Coder/issues/4}}.
Notably, regarding the \textbf{Granite Series~\footnote{\url{https://www.ibm.com/granite/docs/use-cases/prompt-engineering}}} which lack official cross-file templates or examples, we align their processing with the Seed-Coder strategy to maintain the integrity of the orthodox FIM template as is shown in \textbf{Table~\ref{tab:fim_templates}}.

\subsection{Metrics}\label{section:metrics}

To comprehensively evaluate our models, two groups of metrics are adopted: \textbf{Accuracy Metrics} and \textbf{Experience Driven Metrics}, as detailed in \textbf{Table~\ref{tab:evaluation_metrics}}. For accuracy metrics, we tested Pass@1 on Santacoder-fim-task, and measured EM and ES on all language subsets of aiXcoder-FIM-Evaluation and CrossCodeEval described above. For benchmark tables that report language-specific columns and an Overall column, Overall is computed based on all benchmark instances by weighting each language subset according to its number of examples. For experience-driven metrics, we calculated the prefix/suffix code-echo ratios across all models and measured the average generation length (in tokens) for the \Coder\ series models on all the benchmarks.

\begin{table}[htbp]
  \centering
  \begin{threeparttable}
  \caption{Details of Evaluation Metrics.}
  \label{tab:evaluation_metrics}
  \small
  \setlength{\tabcolsep}{4pt}
  \begin{tabular}{lcccccc}
    \toprule

    \multirow{5}{*}{\textbf{Metric}} & \multicolumn{3}{c}{\textbf{Precision}} & \multicolumn{3}{c}{\textbf{Experience Driven}} \\

    \cmidrule(lr){2-4} \cmidrule(lr){5-7}

    & \makecell{\textbf{Exact} \\ \textbf{Match} \\ \textbf{(EM)}} & \makecell{\textbf{Edit} \\ \textbf{Similarity} \\ \textbf{(ES)}} & \makecell{\textbf{Pass@1}} & \textbf{\makecell{Suffix \\ Code-echo \\ Ratio}} & \textbf{\makecell{Prefix \\ Code-echo \\ Ratio}} & \textbf{\makecell{Mean \\ Generation \\ Length}} \\

    \midrule

    \makecell{\textbf{Lower} \\ \textbf{Bound}} & 0 & 0 & 0 & 0 & 0 & 0 \\

    \makecell{\textbf{Upper} \\ \textbf{Bound}} & 100 & 100 & 100 & 100 & 100 & \makecell{\textbf{Benchmark} \\ \textbf{Limit}\tnote{*}} \\

    \makecell{\textbf{Desired} \\ \textbf{Trend}} & $\uparrow$ & $\uparrow$ & $\uparrow$ & $\downarrow$ & $\downarrow$ & $\downarrow$ \\

    \bottomrule
  \end{tabular}
  \begin{tablenotes}
    \item[*] For Mean Generation Length, the upper bound is the benchmark-specific maximum generation length used during inference: 64 generated tokens for Santacoder-fim-task, 256 for aiXcoder-FIM-Evaluation, and 50 for CrossCodeEval.
  \end{tablenotes}
  \end{threeparttable}
\end{table}

\subsubsection{Precision Metrics}

This metric group assesses the correctness of the generated completion. For all of these metrics, a higher score indicates better performance ($\uparrow$ in \textbf{Table~\ref{tab:evaluation_metrics}}).

\begin{itemize}
    \item {\textbf{EM}}, which measures the percentage of predictions that perfectly match the ground truth.

    \item {\textbf{ES}}, which provides a more nuanced score based on the similarity between the prediction and the reference, calculated with \texttt{fuzz.ratio()} in \textit{FuzzyWuzzy} library~\footnote{\url{https://github.com/seatgeek/fuzzywuzzy}}.

    \item {\textbf{Pass@1}} is a functional correctness metric that evaluates if the generated code passes \textbf{all} unit tests on single attempt after being combined with corresponding prefix and suffix to be complete code file.
\end{itemize}

\subsubsection{Experience-Driven Metrics}

We use Experience-Driven Metrics to capture developer-facing efficiency and usability aspects that are not fully reflected by accuracy metrics. These metrics are interpreted together with EM, ES, and Pass@1; lower values indicate fewer erroneous context echoes or shorter outputs under comparable accuracy.

\paragraph{\textbf{Code-echo Ratio}} measures the percentage of completions that start by erroneously repeating an adjacent context line. Because a correct completion may legitimately repeat nearby code, we count code-echo only when the generated first non-empty line differs from the first non-empty line of the canonical solution. All comparisons are conducted after stripping leading and trailing whitespace, and the judgment is restricted to the first non-empty generated line and the adjacent context lines. This operational definition follows the representative adjacent-context echo cases analyzed for \textbf{AP7}/\textbf{STG5}: it targets completions that begin by copying the neighboring prefix or suffix line, while more complex cross-line or long-span repetition would require separate detection rules. We therefore distinguish two directional sub-patterns:
\begin{itemize}
    \item We flag a \textbf{Prefix Code-echo} when an incorrect completion's first non-empty line is identical to the last non-empty line of the prefix while different from the first non-empty line of the canonical solution.
    \item We flag a \textbf{Suffix Code-echo} when an incorrect completion's first non-empty line is identical to the first non-empty line of the suffix while different from the first non-empty line of the canonical solution.
\end{itemize}
We report the \textbf{Prefix Code-echo Ratio} and \textbf{Suffix Code-echo Ratio} as the percentage of all completions that satisfy each directional condition.

\paragraph{\textbf{Mean Generation Length}} is the average number of generated tokens for a model on a benchmark. We use it as an efficiency-oriented metric under the same inference setting and interpret it only together with accuracy metrics. Shorter output is not inherently better, since an empty or incomplete completion is useless; however, when precision is comparable or higher, a lower mean generation length indicates more concise completions and lower inference cost.
Note that differences in tokenizers across models can render direct metric comparisons between different model families meaningless. To ensure the validity of our evaluation, the scope of comparison of mean generation length is restricted to models that share the same base model, as is classified in \textbf{Table~\ref{tab:model_comparison}}.

\section{Results}\label{section:results}

From both perspectives of precision and  experience as is described in \textbf{Section~\ref{section:metrics}}, we elaborate the evaluation results in this section. To conserve space in data tables, \textit{DeepSeek} and \textit{JavaScript} are abbreviated as \textit{DS} and \textit{JS}, respectively.

\subsection{Precision Metrics}\label{section:precision-metrics-results}

\subsubsection{Test-based Metric}\label{section:test-based-metric-results}

Pass@1, a test-based precision metric, is calculated only on Santacoder-fim-task, the benchmark with executable test cases in our evaluation.
For the inference process,
we employed deterministic decoding ($\texttt{temperature}=0$) with the following hyperparameters: $\texttt{top\_p}=1.0$, $\texttt{frequency\_penalty}=0.0$, $\texttt{repetition\_penalty}=1.0$, and $\texttt{length\_penalty}=1.0$. The maximum sequence length is 2048 tokens and the max generation length is capped at 64 tokens, while no other truncation strategies were applied. This specific setup is chosen to evaluate the model's ability to terminate its output at the appropriate time in a single-line completion benchmark like Santacoder-fim-task.

\begin{table}[htb]
\centering
\scriptsize
\caption{Performance of Models on the Santacoder-fim-task Dataset~\citep{allal2023santacoderdontreachstars}.}
\label{tab:model_performance_santacoder}
\setlength{\tabcolsep}{8pt}
\begin{tabular}{l>{\centering\arraybackslash}p{1cm}>{\centering\arraybackslash}p{1cm}>{\centering\arraybackslash}p{1cm}>{\centering\arraybackslash}p{1cm}}
\toprule

\multirow{2}{*}{\textbf{Model}} & \multicolumn{1}{c}{\textbf{JS}} & \multicolumn{1}{c}{\textbf{Java}} & \multicolumn{1}{c}{\textbf{Python}} & \multicolumn{1}{c}{\textbf{Overall}} \\

 & \textbf{Pass@1} & \textbf{Pass@1} & \textbf{Pass@1} & \textbf{Pass@1} \\

\midrule

aiXcoder-7B & 55.1 & 57.8 & 53.0 & 55.6 \\

aiXcoder-7B-v2 & 93.0 & 90.7 & 86.8 & 90.8 \\

\midrule

CodeGemma-7B & 92.5 & 77.6 & 88.6 & 86.1 \\

\midrule

Granite-3.3-8B-Base & 67.4 & 57.1 & 64.0 & 62.8 \\

\midrule

StarCoder2-7B & 20.0 & 16.4 & 1.7 & 14.7 \\

\midrule

DS-Coder-6.7B-Base & 92.0 & 93.4 & 90.8 & 92.3 \\

{DS-Coder-V2-Lite-Base} & 94.0 & 88.9 & 87.6 & 90.7 \\

\midrule

Qwen2.5-Coder-7B & 81.7 & 86.8 & 85.7 & 84.5 \\

\midrule

\SCQwen-SFT-wo-H & \makecell{{\textit{97.2}}} & \makecell{{\textit{96.9}}} & \makecell{{\textit{95.1}}} & \makecell{{\textit{96.6}}} \\

\SCQwen-SFT & \makecell{{\textit{97.3}}} & \makecell{{\textit{97.6}}} & \makecell{{{\textit{95.8}}}} & \makecell{{\textit{97.1}}} \\

\SCQwen-SFT-S & \makecell{{\textit{97.2}}} & \makecell{{\textit{97.6}}} & \makecell{{{\textit{95.8}}}} & \makecell{{\textit{97.0}}} \\

\SCQwen-DPO-wo-RS-S & \makecell{{\textit{97.3}}} & \makecell{{\textit{97.6}}} & \makecell{{\textit{95.6}}} & \makecell{{\textit{97.0}}} \\

\textbf{\SCQwen-DPO-S} & \makecell{{\textbf{\textit{97.4}}}} & \makecell{{\textbf{\textit{97.7}}}} & \makecell{{\textbf{\textit{95.8}}}} & \makecell{{\textbf{\textit{97.2}}}} \\

\midrule

Seed-Coder-8B-Base & 92.6 & 91.4 & 90.0 & 91.6 \\

\midrule

\SCSeed-SFT-wo-H & \makecell{{\textit{95.5}} } & \makecell{{\textit{88.0}} } & \makecell{{\textit{90.3}} } & \makecell{{\textit{91.6}} } \\

\SCSeed-SFT & \makecell{{\textit{93.8}} } & \makecell{{\textit{93.8}} } & \makecell{{\textit{90.1}} } & \makecell{{\textit{93.0}} } \\

\SCSeed-DPO-wo-RS & \makecell{{\textit{93.7}} } & \makecell{{\textit{93.9}} } & \makecell{{\textit{90.1}} } & \makecell{{\textit{93.0}} } \\

\textbf{\SCSeed-DPO} & \textit{95.5} & \textit{94.0} & \textit{90.3} & \textit{93.8} \\

\bottomrule
\end{tabular}
\end{table}

Results of the Santacoder-fim-task benchmark are summarized in \textbf{Table~\ref{tab:model_performance_santacoder}}.
Under the Overall computation protocol described in \textbf{Section~\ref{section:metrics}}, \CoderQwen-DPO-S achieves the highest overall Pass@1 score among the evaluated models, with $97.2$, while \CoderSeed-DPO delivers robust performance with a score of $93.8$. The Qwen-based variants after post-training all reach the high-90\% range, indicating that Santacoder-fim-task is approaching saturation for strong in-file FIM models.

On this benchmark, the SFT stage accounts for most of the visible improvement, especially for the Qwen family: Qwen2.5-Coder-7B starts at $84.5$, \CoderQwen-SFT reaches $97.1$, and the retained Seed-filtered \CoderQwen-SFT-S remains close at $97.0$. For the \CoderSeed\ family, \CoderSeed-SFT improves from the Seed-Coder-8B-Base baseline of $91.6$ to $93.0$, and \CoderSeed-DPO reaches $93.8$. Since the post-trained Qwen variants are already in the high-90\% range, we use Santacoder-fim-task mainly to describe the short in-file completion progression, while relying on broader text-based benchmarks for more discriminative ablation evidence.

Residual inspection supports this interpretation without turning Santacoder-fim-task into the main diagnostic focus. The canonical completions pass under the same fully provisioned evaluation setup, while the remaining failures are mostly wrong-answer cases involving fine-grained semantic constraints, including ordering constraints, with a smaller number of compilation, syntax/indentation, and timeout failures caused by non-terminating logic. These cases suggest that the residual errors concentrate on boundary or language-specific details rather than missing evaluation dependencies, so broader benchmarks provide a more discriminative view.

Therefore, we do not interpret Santacoder-fim-task as the sole differentiator for modern FIM models once performance reaches the high-90\% range. Instead, we use it primarily as a short in-file completion benchmark and rely on the broader benchmark suite for a more discriminative view of execution-level capability.

\subsubsection{Text-based Metric}\label{section:text-based-metric-results}

EM and ES, two text-based precision metrics, are calculated in  aiXcoder-FIM-Evaluation and CrossCodeEval benchmark.

For the model inference process in {aiXcoder-FIM-Evaluation} benchmark, we set the input context to 8192 tokens and cap the maximum output length at 256 tokens, using deterministic decoding ($\texttt{temperature}=0$) with the following hyperparameters: $\texttt{top\_p}=1.0$, $\texttt{frequency\_penalty}=0.0$, $\texttt{repetition\_penalty}=1.0$, and $\texttt{length\_penalty}=1.0$.

With regard to the {CrossCodeEval} benchmark, we set the input context to 8192 tokens and the cross-file context length to 2048 tokens while capping the maximum output length at 50 tokens, using deterministic decoding ($\texttt{temperature}=0$) with the following hyperparameters: $\texttt{top\_p}=1.0$, $\texttt{frequency\_penalty}=0.0$, $\texttt{repetition\_penalty}=1.0$, and $\texttt{length\_penalty}=1.0$.

\begin{table}[htbp]
\centering
\setlength{\tabcolsep}{0.01pt}
\scriptsize
\caption{Performance of Models on the aiXcoder-FIM-Evaluation Dataset~\citep{jiang2024aixcoder}.}
\label{tab:model_performance_aixcoder}
\begin{tabular}{l>{\centering\arraybackslash}p{0.77cm}>{\centering\arraybackslash}p{0.77cm}>{\centering\arraybackslash}p{0.77cm}>{\centering\arraybackslash}p{0.77cm}>{\centering\arraybackslash}p{0.77cm}>{\centering\arraybackslash}p{0.77cm}>{\centering\arraybackslash}p{0.77cm}>{\centering\arraybackslash}p{0.77cm}>{\centering\arraybackslash}p{0.77cm}>{\centering\arraybackslash}p{0.77cm}}
\toprule

\multirow{2}{*}{\textbf{Model}} & \multicolumn{2}{c}{\textbf{C++}} & \multicolumn{2}{c}{\textbf{JS}} & \multicolumn{2}{c}{\textbf{Java}} & \multicolumn{2}{c}{\textbf{Python}} & \multicolumn{2}{c}{\textbf{Overall}} \\

\cmidrule(lr){2-3} \cmidrule(lr){4-5} \cmidrule(lr){6-7} \cmidrule(lr){8-9} \cmidrule(lr){10-11}

 & \textbf{EM} & \textbf{ES} & \textbf{EM} & \textbf{ES} & \textbf{EM} & \textbf{ES} & \textbf{EM} & \textbf{ES} & \textbf{EM} & \textbf{ES} \\

\midrule

aiXcoder-7B & 40.3 & 71.4 & 42.4 & 72.2 & 57.5 & 82.2 & 39.6 & 72.0 & 45.2 & 74.6 \\

aiXcoder-7B-v2 & 48.3 & 77.3 & 46.1 & 75.4 & 57.9 & 83.0 & 41.9 & 73.1 & 48.6 & 77.3 \\

\midrule

CodeGemma-7B & 47.0 & 76.0 & 44.9 & 74.8 & 56.7 & 83.0 & 46.5 & 76.6 & 48.9 & 77.7 \\

\midrule

Granite-3.3-8B-Base & 32.2 & 62.2 & 29.2 & 61.3 & 38.7 & 67.7 & 33.3 & 63.8 & 33.4 & 63.8 \\

\midrule

StarCoder2-7B & 14.2 & 45.2 & 6.8 & 44.9 & 4.8 & 40.1 & 20.4 & 51.2 & 11.5 & 45.3 \\

\midrule

DS-Coder-6.7B-Base & 48.8 & 75.2 & 46.1 & 75.3 & 59.2 & 83.0 & 47.7 & 76.7 & 50.6 & 77.7 \\

{DS-Coder-V2-Lite-Base} & 49.8 & 76.3 & 47.0 & 75.3 & 59.5 & 83.0 & 47.2 & 75.9 & 51.0 & 77.7 \\

\midrule

Qwen2.5-Coder-7B & 48.9 & 73.7 & 47.8 & 74.5 & 58.1 & 80.3 & 47.8 & 76.1 & 50.8 & 76.2 \\

\midrule

\SCQwen-SFT-wo-H & \makecell{{\textit{54.2}}} & \makecell{{\textit{82.0}}} & \makecell{{\textit{53.1}}} & \makecell{{\textit{81.0}}} & \makecell{{\textit{64.0}}} & \makecell{{{\textit{87.7}}}} & \makecell{{\textit{51.2}}} & \makecell{{\textit{80.2}}} & \makecell{{\textit{55.6}}} & \makecell{{\textit{82.7}}} \\

\SCQwen-SFT & \makecell{{\textit{54.6}}} & \makecell{{{\textit{82.4}}}} & \makecell{{\textit{53.6}}} & \makecell{{\textit{81.4}}} & \makecell{{\textit{64.0}}} & \makecell{{\textit{87.7}}} & \makecell{{\textit{51.4}}} & \makecell{{\textit{80.3}}} & \makecell{{\textit{55.9}}} & \makecell{{\textit{83.0}}} \\

\SCQwen-SFT-S & \makecell{{\textit{54.4}}} & \makecell{{{\textit{82.3}}}} & \makecell{{\textit{53.4}}} & \makecell{{\textit{81.2}}} & \makecell{{\textit{64.0}}} & \makecell{{\textit{87.7}}} & \makecell{{\textit{51.4}}} & \makecell{{\textit{80.2}}} & \makecell{{\textit{55.9}}} & \makecell{{\textit{82.9}}} \\

\SCQwen-DPO-wo-RS-S & \makecell{{\textit{54.4}}} & \makecell{{\textit{82.1}}} & \makecell{{\textit{53.7}}} & \makecell{{{\textit{81.3}}}} & \makecell{{\textit{63.9}}} & \makecell{{\textit{87.7}}} & \makecell{{{\textit{51.4}}}} & \makecell{{\textit{80.3}}} & \makecell{{\textit{56.0}}} & \makecell{{\textit{82.9}}} \\

\textbf{\SCQwen-DPO-S} & \makecell{{\textbf{\textit{54.8}}}} & \makecell{\textbf{\textit{82.4}}} & \makecell{{\textbf{\textit{53.8}}}} & \makecell{\textbf{\textit{81.4}}} & \makecell{{\textbf{\textit{64.1}}}} & \makecell{\textbf{\textit{87.8}}} & \makecell{\textbf{\textit{51.4}}} & \makecell{\textit{80.4}} & \makecell{{\textbf{\textit{56.1}}}} & \makecell{{\textbf{\textit{83.0}}}} \\

\midrule

Seed-Coder-8B-Base & 48.7 & 75.3 & 43.1 & 70.0 & 53.8 & 76.2 & 47.4 & 75.3 & 48.3 & 74.2 \\

\midrule

\SCSeed-SFT-wo-H & \textit{49.6} & \textit{76.6} & \textit{45.8} & \textit{73.8} & \textit{57.9} & \textit{81.3} & \textit{48.4} & \textit{76.8} & \textit{50.5} & \textit{77.2} \\

\SCSeed-SFT & \textit{51.4} & \textit{78.1} & \textit{49.2} & \textit{77.5} & \textit{60.3} & \textit{83.8} & \textit{49.7} & \textit{78.4} & \textit{52.8} & \textit{79.6} \\

\SCSeed-DPO-wo-RS & \textit{53.8} & \textit{81.3} & \textit{50.5} & \textit{79.0} & \textit{63.3} & \textit{87.0} & \textit{50.7} & \textit{79.8} & \textit{54.7} & \textit{81.8} \\

\textbf{\SCSeed-DPO} & \textit{54.2} & \textit{82.0} & \textit{50.6} & \textit{79.4} & \textit{63.5} & \textit{87.2} & \textit{51.2} & \textbf{\textit{80.6}} & \textit{55.0} & \textit{82.4} \\

\bottomrule
\end{tabular}
\end{table}

The results of the aiXcoder-FIM-Evaluation are summarized in \textbf{Table~\ref{tab:model_performance_aixcoder}}. Among the evaluated models, \CoderQwen-DPO-S obtains the highest overall scores, with 56.1 EM and 83.0 ES, exceeding DeepSeek-Coder-V2-Lite-Base (51.0 EM). \CoderSeed-DPO also improves over its base model, reaching 55.0 EM and 82.4 ES compared with 48.3 EM and 74.2 ES for Seed-Coder-8B-Base. These results indicate that the proposed post-training pipeline improves both model families on in-file text-based completion.

Ablation study investigates the incremental contributions of each training stage in aiXcoder-FIM-Evaluation. For the \CoderSeed\ family, the SFT stage yields a substantial boost. Specifically, the inclusion of human behavior data (\CoderSeed-SFT) elevates performance to 52.8 EM, a marked improvement over the version without human data (\CoderSeed-SFT-wo-H) at 50.5 EM. The Qwen family follows the same SFT method with Qwen-specific filtering by default: \CoderQwen-SFT improves from 55.6 EM and 82.7 ES in \CoderQwen-SFT-wo-H to 55.9 EM and 83.0 ES. The retained \CoderQwen-SFT-S row shows that the Seed-filtered SFT chain reaches the same 55.9 EM. The subsequent alignment stage provides the final refinement. For \CoderSeed\ , applying DPO with repetition suppression raises scores from 52.8 to 55.0 EM. The Qwen DPO-stage variants use the Seed-PPL data pipeline to match the Seed DPO variants under the same SFT/DPO input corpora and training procedure while changing the pretrained model family from Seed to Qwen. Within this cross-model-family matched-data setting, the comparison between \CoderQwen-DPO-wo-RS-S and \CoderQwen-DPO-S isolates \textbf{STG5} and reaches a peak of 56.1 EM. This progression indicates that the SFT-stage heuristic masking rules improve both model families, while the matched Qwen DPO comparison supports the role of repetition suppression.

\begin{table}[htbp]
\centering
\setlength{\tabcolsep}{0.01pt}
\scriptsize
\caption{Performance of Models on the CrossCodeEval Dataset~\citep{ding2023crosscodeeval}.}
\label{tab:model_performance_compare_cceval}
\begin{tabular}{l>{\centering\arraybackslash}p{0.77cm}>{\centering\arraybackslash}p{0.77cm}>{\centering\arraybackslash}p{0.77cm}>{\centering\arraybackslash}p{0.77cm}>{\centering\arraybackslash}p{0.77cm}>{\centering\arraybackslash}p{0.77cm}>{\centering\arraybackslash}p{0.77cm}>{\centering\arraybackslash}p{0.77cm}>{\centering\arraybackslash}p{0.77cm}>{\centering\arraybackslash}p{0.77cm}}
\toprule

\multirow{2}{*}{\textbf{Model}} & \multicolumn{2}{c}{\textbf{C\#}} & \multicolumn{2}{c}{\textbf{Java}} & \multicolumn{2}{c}{\textbf{Python}} & \multicolumn{2}{c}{\textbf{TS}} & \multicolumn{2}{c}{\textbf{Overall}} \\\cmidrule(lr){2-3} \cmidrule(lr){4-5} \cmidrule(lr){6-7} \cmidrule(lr){8-9} \cmidrule(lr){10-11} & \textbf{EM} & \textbf{ES} & \textbf{EM} & \textbf{ES} & \textbf{EM} & \textbf{ES} & \textbf{EM} & \textbf{ES} & \textbf{EM} & \textbf{ES} \\

\midrule

aiXcoder-7B & 40.9 & 79.2 & 38.6 & 77.6 & 3.4 & 48.1 & 32.1 & 74.3 & 27.4 & 68.9 \\

aiXcoder-7B-v2 & 63.1 & 90.4 & 52.2 & \textbf{86.0} & 43.0 & 80.0 & 47.7 & 85.6 & 50.1 & 85.0 \\

\midrule

CodeGemma-7B & 6.1 & 63.9 & 2.9 & 57.8 & 3.5 & 57.4 & 2.9 & 65.6 & 3.6 & 61.4 \\

\midrule

Granite-3.3-8B-Base & 34.2 & 77.1 & 25.8 & 70.5 & 29.5 & 72.5 & 24.4 & 69.1 & 27.8 & 71.7 \\

\midrule

StarCoder2-7B & 8.8 & 75.8 & 7.5 & 71.2 & 11.4 & 64.8 & 7.5 & 73.8 & 8.8 & 71.2 \\

\midrule

DS-Coder-6.7B-Base & 61.8 & 88.2 & 51.8 & 84.2 & 46.9 & 80.7 & 50.2 & 84.2 & 51.7 & 84.0 \\

{DS-Coder-V2-Lite-Base} & 59.2 & 86.9 & 50.1 & 83.5 & 45.7 & 79.9 & 47.5 & 83.0 & 49.7 & 83.0 \\

\midrule

Qwen2.5-Coder-7B & 61.5 & 87.8 & 50.1 & 83.7 & 45.4 & 79.8 & 47.9 & 83.7 & 50.1 & 83.4 \\

\midrule

\SCQwen-SFT-wo-H & \makecell{{\textit{63.9}}} & \makecell{{\textit{90.1}}} & \makecell{{\textit{47.8}}} & \makecell{{\textit{82.3}}} & \makecell{{\textit{45.7}}} & \makecell{{\textit{80.7}}} & \makecell{{\textit{48.3}}} & \makecell{{\textit{85.1}}} & \makecell{{\textit{50.3}}} & \makecell{{\textit{84.2}}} \\

\SCQwen-SFT & \makecell{{\textit{64.2}}} & \makecell{{\textit{90.5}}} & \makecell{{\textit{50.1}}} & \makecell{{\textit{84.3}}} & \makecell{{\textit{46.0}}} & \makecell{{\textit{81.0}}} & \makecell{{\textit{49.2}}} & \makecell{{\textit{85.3}}} & \makecell{{\textit{51.2}}} & \makecell{{\textit{84.9}}} \\

\SCQwen-SFT-S & \makecell{{\textit{62.6}}} & \makecell{{\textit{89.8}}} & \makecell{{\textit{49.8}}} & \makecell{{\textit{84.1}}} & \makecell{{\textit{45.8}}} & \makecell{{\textit{80.5}}} & \makecell{{\textit{47.7}}} & \makecell{{\textit{84.5}}} & \makecell{{\textit{50.3}}} & \makecell{{\textit{84.3}}} \\

\SCQwen-DPO-wo-RS-S & \makecell{{\textit{59.3}}} & \makecell{{\textit{89.6}}} & \makecell{{\textit{50.9}}} & \makecell{{\textit{85.3}}} & \makecell{{\textit{46.4}}} & \makecell{{\textit{81.2}}} & \makecell{{\textit{48.9}}} & \makecell{{\textit{85.3}}} & \makecell{{\textit{50.5}}} & \makecell{{\textit{85.0}}} \\

\textbf{\SCQwen-DPO-S} & \makecell{{\textit{64.4}}} & \makecell{{\textit{90.1}}} & \makecell{{\textit{51.6}}} & \makecell{{\textit{85.5}}} & \makecell{{\textit{46.6}}} & \makecell{{\textit{81.4}}} & \makecell{{\textit{50.7}}} & \makecell{{\textit{85.8}}} & \makecell{{\textit{52.2}}} & \makecell{{\textit{85.3}}} \\

\midrule

Seed-Coder-8B-Base & \makecell{{61.9}} & \makecell{{89.1}} & \makecell{{52.6}} & \makecell{{84.5}} & \makecell{{49.3}} & \makecell{{81.9}} & \makecell{{50.6}} & \makecell{{84.8}} & \makecell{{52.7}} & \makecell{{84.7}} \\

\midrule

\SCSeed-SFT-wo-H & \makecell{{\textit{62.3}}} & \makecell{{\textit{89.5}}} & \makecell{{\textit{51.3}}} & \makecell{{\textit{84.5}}} & \makecell{{\textit{49.8}}} & \makecell{{\textit{82.3}}} & \makecell{{\textit{50.8}}} & \makecell{{\textit{85.3}}} & \makecell{{\textit{52.7}}} & \makecell{{\textit{85.2}}} \\

\SCSeed-SFT & \makecell{{\textit{63.5}}} & \makecell{{\textit{89.8}}} & \makecell{{\textit{52.0}}} & \makecell{{\textit{84.4}}} & \makecell{{\textit{49.8}}} & \makecell{{\textit{82.4}}} & \makecell{{\textit{51.5}}} & \makecell{{\textit{85.5}}} & \makecell{{\textit{53.3}}} & \makecell{{\textit{85.2}}} \\

\SCSeed-DPO-wo-RS & \makecell{{\textit{65.4}}} & \makecell{{\textit{91.1}}} & \makecell{{\textit{52.9}}} & \makecell{{\textit{85.4}}} & \makecell{{\textit{48.7}}} & \makecell{{\textit{82.4}}} & \makecell{{\textit{51.7}}} & \makecell{{\textit{86.1}}} & \makecell{{\textit{53.6}}} & \makecell{{\textit{85.8}}} \\

\textbf{\SCSeed-DPO} & \textit{\textbf{66.6}} & \textit{\textbf{91.1}} & \textit{\textbf{53.0}} & \textit{\textbf{85.8}} & \textit{\textbf{49.8}} & \textit{\textbf{82.7}} & \textit{\textbf{52.6}} & \textit{\textbf{86.6}} & \textit{\textbf{54.3}} & \textit{\textbf{86.2}} \\

\bottomrule
\end{tabular}
\end{table}

\begin{table}[!t]
\centering
\begin{threeparttable}
\scriptsize
\caption{Statistical Significance of Ablation Comparisons on Precision Metrics.}
\label{tab:ablation_significance_precision}
\setlength{\tabcolsep}{1.0pt}
\begin{tabular}{>{\raggedright\arraybackslash}p{3.80cm}
*{3}{>{\centering\arraybackslash}p{0.85cm}>{\centering\arraybackslash}p{1.27cm}}}
\toprule

\makecell[l]{\textbf{Model}\\\textbf{Comparison}} &
\multicolumn{2}{c}{\makecell{\textbf{aiXcoder\tnote{*}} \\ \textbf{EM} \\ \textbf{$N=32{,}000$}}} &
\multicolumn{2}{c}{\makecell{\textbf{CrossCodeEval} \\ \textbf{EM} \\ \textbf{$N=9{,}928$}}} &
\multicolumn{2}{c}{\makecell{\textbf{CrossCodeEval} \\ \textbf{C\# EM} \\ \textbf{$N=1{,}768$}}} \\

\cmidrule(lr){2-3} \cmidrule(lr){4-5} \cmidrule(lr){6-7}

 & \textbf{$\Delta$} & \textbf{$p$} & \textbf{$\Delta$} & \textbf{$p$} & \textbf{$\Delta$} & \textbf{$p$} \\

\midrule
\addlinespace[1pt]

\makecell[l]{Qwen2.5-Coder-7B\\$\to$ \SCQwen-SFT\\$\to$ \SCQwen-SFT-S} & \makecell{\strut\\+5.1\\+5.1} & \makecell{\strut\\\textbf{\textless{}0.001}\\\textbf{\textless{}0.001}} & \makecell{\strut\\+1.1\\+0.2} & \makecell{\strut\\\textbf{\textless{}0.001}\\\textbf{0.022}} & \makecell{\strut\\+2.7\\+1.1} & \makecell{\strut\\\textbf{\textless{}0.001}\\\textbf{0.012}} \\

\makecell[l]{\SCQwen-SFT-wo-H\\$\to$ \SCQwen-SFT} & \makecell{\strut\\+0.3} & \makecell{\strut\\\textbf{0.033}} & \makecell{\strut\\+0.9} & \makecell{\strut\\\textbf{\textless{}0.001}} & \makecell{\strut\\+0.3} & \makecell{\strut\\\textbf{0.039}} \\

\makecell[l]{\SCQwen-SFT-S\\$\to$ \SCQwen-DPO-S} & \makecell{\strut\\+0.2} & \makecell{\strut\\\textbf{0.001}} & \makecell{\strut\\+1.9} & \makecell{\strut\\\textbf{\textless{}0.001}} & \makecell{\strut\\+1.8} & \makecell{\strut\\\textbf{0.001}} \\

\makecell[l]{\SCQwen-DPO-wo-RS-S\\$\to$ \SCQwen-DPO-S} & \makecell{\strut\\+0.1} & \makecell{\strut\\\textbf{0.005}} & \makecell{\strut\\+1.7} & \makecell{\strut\\\textbf{\textless{}0.001}} & \makecell{\strut\\+5.1} & \makecell{\strut\\\textbf{\textless{}0.001}} \\

\midrule
\addlinespace[1pt]

\makecell[l]{Seed-Coder-8B-Base\\$\to$ \SCSeed-SFT} & \makecell{\strut\\+4.5} & \makecell{\strut\\\textbf{\textless{}0.001}} & \makecell{\strut\\+0.6} & \makecell{\strut\\\textbf{0.011}} & \makecell{\strut\\+1.6} & \makecell{\strut\\\textbf{0.018}} \\

\makecell[l]{\SCSeed-SFT-wo-H\\$\to$ \SCSeed-SFT} & \makecell{\strut\\+2.3} & \makecell{\strut\\\textbf{\textless{}0.001}} & \makecell{\strut\\+0.6} & \makecell{\strut\\\textbf{0.007}} & \makecell{\strut\\+1.2} & \makecell{\strut\\\textbf{0.045}} \\

\makecell[l]{\SCSeed-SFT\\$\to$ \SCSeed-DPO} & \makecell{\strut\\+2.2} & \makecell{\strut\\\textbf{\textless{}0.001}} & \makecell{\strut\\+1.0} & \makecell{\strut\\\textbf{\textless{}0.001}} & \makecell{\strut\\+3.1} & \makecell{\strut\\\textbf{\textless{}0.001}} \\

\makecell[l]{\SCSeed-DPO-wo-RS\\$\to$ \SCSeed-DPO} & \makecell{\strut\\+0.3} & \makecell{\strut\\\textbf{0.042}} & \makecell{\strut\\+0.7} & \makecell{\strut\\\textbf{0.005}} & \makecell{\strut\\+1.2} & \makecell{\strut\\\textbf{0.039}} \\

\bottomrule
\end{tabular}
\begin{tablenotes}
\item[*] \textbf{aiXcoder}: aiXcoder-FIM-Evaluation.
\item[] Exact one-sided McNemar tests are applied to paired binary EM correctness changes~\cite{Dietterich1998StatisticalTests,Raschka2018ModelEvaluation}. $\Delta$ denotes the arrowed target model's mean metric value minus the first model's value in the same comparison cell, and $p$ denotes the exact one-sided McNemar p-value in the improvement direction.
The \texttt{-S} suffix marks Qwen variants using the Seed-PPL data pipeline.
\end{tablenotes}
\end{threeparttable}
\end{table}

\textbf{Table~\ref{tab:model_performance_compare_cceval}} lists the performance on CrossCodeEval, a diverse multilingual benchmark for cross-file completion~\citep{ding2023crosscodeeval}. Our variants show the best overall performance among the evaluated models. \CoderSeed-DPO leads with 54.3 EM and 86.2 ES, surpassing DS-Coder-6.7B-Base (51.7 EM, 84.0 ES) and its own base model (52.7 EM, 84.7 ES). \CoderQwen-DPO-S also improves over the Qwen base model under the matched Seed-PPL DPO setting, reaching 52.2 EM and 85.3 ES compared with 50.1 EM and 83.4 ES.

The Qwen SFT-stage results use Qwen-specific PPL filtering, and adding \textbf{STG2-STG4} improves the visible overall comparison from 50.3 EM and 84.2 ES in \CoderQwen-SFT-wo-H to 51.2 EM and 84.9 ES in \CoderQwen-SFT.

The retained \CoderQwen-SFT-S row records the Seed-filtered SFT starting point used for the Qwen DPO variants. The Qwen DPO-stage variants use the Seed-PPL data pipeline to match the Seed DPO variants under the same SFT/DPO input corpora and training procedure while changing the pretrained model family from Seed to Qwen. Within this cross-model-family matched-data setting, the comparison between \CoderQwen-DPO-wo-RS-S and \CoderQwen-DPO-S further isolates \textbf{STG5}: removing it drops Qwen from 52.2 EM, 85.3 ES to 50.5 EM, 85.0 ES, while the analogous Seed comparison drops from 54.3 EM, 86.2 ES to 53.6 EM, 85.8 ES.

The significance results provide paired-sample support for the main ablation decisions. For the Qwen family, the Qwen-filtered SFT-stage comparison supports the contribution of \textbf{STG2-STG4}, while the retained \CoderQwen-SFT-S to \CoderQwen-DPO-S row provides overall matched-data alignment evidence under the Seed-PPL data pipeline. The direct \CoderQwen-DPO-wo-RS-S to \CoderQwen-DPO-S comparison further supports \textbf{STG5}.

For the Seed family, adding \textbf{STG2-STG4} and the full alignment stage yields positive EM changes with one-sided McNemar support across the reported EM views. The direct \CoderSeed-DPO-wo-RS to \CoderSeed-DPO comparison further supports \textbf{STG5}, matching the Qwen-side pattern while using the Seed backbone and Seed-PPL data pipeline.

\subsection{Experience Driven Metrics}

Based on the model outputs from the same three benchmarks, we compute the experience-driven metrics outlined in \textbf{Table~\ref{tab:evaluation_metrics}}.

\subsubsection{Prefix and Suffix Code-echo Ratios}\label{section:suffix-prefix-repeat}

The prevalence of erroneous prefix and suffix code-echo in the model outputs is analyzed with the results presented in \textbf{Table~\ref{tab:model_performance_rep_rate}}. Admittedly, inference parameters may affect the repetition rate. We make sure that all the models are evaluated using the same settings as is described in generation.

\begin{table}[htbp]
\centering
\begin{threeparttable}
\scriptsize
\caption{Prefix and Suffix Code-Echo Ratio by Model and Dataset.}
\label{tab:model_performance_rep_rate}
\setlength{\tabcolsep}{0.5pt}
\begin{tabular}{l>{\centering\arraybackslash}p{0.9cm}>{\centering\arraybackslash}p{0.9cm}>{\centering\arraybackslash}p{0.9cm}>{\centering\arraybackslash}p{0.9cm}>{\centering\arraybackslash}p{0.9cm}>{\centering\arraybackslash}p{0.9cm}>{\centering\arraybackslash}p{1.0cm}>{\centering\arraybackslash}p{1.0cm}}
\toprule

\multirow{2}{*}{\textbf{Model}} & \multicolumn{2}{c}{\textbf{aiXcoder\tnote{*}}} & \multicolumn{2}{c}{\textbf{Santacoder\tnote{\dag}}} & \multicolumn{2}{c}{\textbf{CCE\tnote{\S}}} & \multicolumn{2}{c}{\textbf{C\# in CCE\tnote{\S}}} \\

\cmidrule(lr){2-3} \cmidrule(lr){4-5} \cmidrule(lr){6-7} \cmidrule(lr){8-9}

 & \textbf{Prefix} & \textbf{Suffix} & \textbf{Prefix} & \textbf{Suffix} & \textbf{Prefix} & \textbf{Suffix} & \textbf{Prefix} & \textbf{Suffix} \\

\midrule

aiXcoder-7B & 0.09 & 2.83 & \textbf{0.00} & 0.35    & 0.05 & 5.51 & 0.17 & 0.68  \\

aiXcoder-7B-v2 & 0.04 & 2.27 & \textbf{0.00} & 0.29 & 0.00 & 2.22 & 0.00 & 0.00  \\

\midrule

CodeGemma-7B & 0.05 & 0.82 & \textbf{0.00} & 0.04     & 0.01 & 0.30 & 0.00 & 0.00  \\

\midrule

Granite-3.3-8B & 0.18 & 2.98 & \textbf{0.00} & 1.15      & 0.10 & 1.31 & 0.00 & 0.06  \\

\midrule

StarCoder2-7B & 0.13 & 1.59 & \textbf{0.00} & 0.21  & 0.00 & 0.26 & 0.00 & 0.00  \\

\midrule

DS-Coder-6.7B-Base & 0.09 & 0.92 & \textbf{0.00} & \textbf{0.00}     & \textbf{0.00} & 0.33 & \textbf{0.00} & \textbf{0.00}  \\

{DS-Coder-V2-Lite-Base} & 0.12 & 0.88 & \textbf{0.00} & 0.04  & \textbf{0.00} & 0.53 & \textbf{0.00} & \textbf{0.00}  \\

\midrule

Qwen2.5-Coder-7B & 0.09 & 3.47 & \textbf{0.00} & 0.08  & \textbf{0.00} & 0.44 & \textbf{0.00} & \textbf{0.00}  \\

\midrule

\SCQwen-SFT-wo-H & \textit{0.04} & \textit{0.32} & \textit{\textbf{0.00}} & \textit{0.08} & \textit{\textbf{0.00}} & \textit{0.18} & \textit{\textbf{0.00}} & \textit{\textbf{0.00}}  \\

\SCQwen-SFT & \textit{0.04} & \textit{0.30} & \textit{\textbf{0.00}} & \textit{0.07} & \textit{\textbf{0.00}} & \textit{0.18} & \textit{\textbf{0.00}} & \textit{\textbf{0.00}}  \\

\SCQwen-SFT-S & \textit{0.05} & \textit{0.29} & \textit{\textbf{0.00}} & \textit{0.15}         & \textit{\textbf{0.00}} & \textit{0.19} & \textit{\textbf{0.00}} & \textit{\textbf{0.00}}  \\

\SCQwen-DPO-wo-RS-S & \textit{0.04} & \textit{0.22} & \textit{\textbf{0.00}} & \textit{0.10}   & \textit{\textbf{0.00}} & \textit{0.19} & \textit{\textbf{0.00}} & \textit{\textbf{0.00}}  \\

\textbf{\SCQwen-DPO-S} & \textit{\textbf{0.03}} & \textit{\textbf{0.19}} & \textit{\textbf{0.00}} & \textit{0.06} & \textit{\textbf{0.00}} & \textit{\textbf{0.14}} & \textit{\textbf{0.00}} & \textit{\textbf{0.00}}  \\

\midrule

Seed-Coder-8B-Base & 0.24 & 1.47 & \textbf{0.00} & 0.08  & 0.00 & 0.50 & 0.00 & 0.06  \\

\midrule

\SCSeed-SFT-wo-H & \textit{0.14} & \textit{1.28} & \textit{\textbf{0.00}} & \textit{0.10}   & \textit{\textbf{0.00}} & \textit{0.30} & \textit{\textbf{0.00}} & \textit{\textbf{0.00}}  \\

\SCSeed-SFT & \textit{0.10} & \textit{0.77} & \textit{\textbf{0.00}} & \textit{0.10}        & \textit{\textbf{0.00}} & \textit{0.39} & \textit{\textbf{0.00}} & \textit{\textbf{0.00}}  \\

\SCSeed-DPO-wo-RS & \textit{0.05} & \textit{0.76} & \textit{\textbf{0.00}} & \textit{0.08}  & \textit{\textbf{0.00}} & \textit{0.40} & \textit{\textbf{0.00}} & \textit{\textbf{0.00}}  \\

\textbf{\SCSeed-DPO} & \textit{0.04} & \textit{0.60} & \textit{\textbf{0.00}} & \textit{0.06}  & \textit{\textbf{0.00}} & \textit{0.23} & \textit{\textbf{0.00}} & \textit{\textbf{0.00}}  \\

\bottomrule
\end{tabular}
\begin{tablenotes}
\item[*] \textbf{aiXcoder}: aiXcoder-FIM-Evaluation;
\item[\dag] \textbf{Santacoder}: Santacoder-fim-task;
\item[\S] \textbf{CCE}: CrossCodeEval.
\item All reported ratios are percentages (\%); the \% symbol is omitted for brevity.
\end{tablenotes}
\end{threeparttable}
\end{table}

A primary observation is that {suffix code-echo is far more prevalent than Prefix code-echo}, regardless of the model's preference on PSM or SPM mode. On the aiXcoder-FIM-Evaluation benchmark, suffix ratios are often an order of magnitude higher, while on the Santacoder-fim-task benchmark, prefix code-echo is entirely absent (0.00\%) for all models.

Our analysis of the training stages across both \CoderSeed\ and Qwen families shows that our fine-tuning reduces these code-echo errors. For instance, in the \CoderSeed\ series, \CoderSeed-SFT reduces the suffix ratio to 0.77\% on aiXcoder-FIM-Evaluation (compared to 1.47\% in Seed-Coder-8B-Base) and maintains 0.10\% on Santacoder-fim-task. For the Qwen-specific SFT chain, the suffix echo ratio on aiXcoder-FIM-Evaluation drops from 3.47\% in Qwen2.5-Coder-7B-Base to 0.32\% in \CoderQwen-SFT-wo-H and 0.30\% in \CoderQwen-SFT. The retained Seed-PPL Qwen SFT anchor, \CoderQwen-SFT-S, remains close at 0.29\% and serves as the starting point for the matched-data Qwen DPO variants.

The subsequent alignment stage provides further improvements, as seen in the intermediate \CoderQwen-DPO-wo-RS-S and \CoderSeed-DPO-wo-RS variants, even without specialized suppression data. One plausible reason is that general rejected samples in the preference data naturally include code-echo examples (\textbf{Figure~\ref{fig:rejected-2}}). Adding repetition-suppression data in \textbf{STG5} further reduces echo ratios. \CoderSeed-DPO reaches a suffix ratio of 0.60\% on aiXcoder-FIM-Evaluation, 0.06\% in Santacoder-fim-task, and 0.23\% on CrossCodeEval. The Qwen DPO rows use the same Seed-PPL data pipeline as the Seed DPO rows; within this cross-model-family matched-data setting, \CoderQwen-DPO-S reaches 0.19\%, 0.06\%, and 0.14\%, respectively.

\subsubsection{Mean Generation Length}\label{section:mean-generation-length}

As previously discussed, differences in tokenization across model series imply that comparisons of output length are only meaningful within the same model series. Therefore, we compare the average length of code completion outputs solely among the Seed-Coder and Qwen2.5-Coder series of models as is shown in \textbf{Table~\ref{tab:model_performance_gen_length}}.

The results show that \CoderSeed-DPO and \CoderQwen-DPO-S yield the shortest generation lengths within their respective model series. These models also achieve the highest accuracy within the variants in their respective model families across the benchmarks discussed in \textbf{Section~\ref{section:precision-metrics-results}}, suggesting that the shorter outputs do not come from simply under-generating code.

The fine-tuning stage initiates this reduction.
For the \CoderSeed\ series, the inclusion of human behavior data in \textbf{STG2-STG4} reduces average tokens from 68.0 to 58.7 on aiXcoder-FIM-Evaluation and from 22.1 to 17.2 on CrossCodeEval when compared with \CoderSeed-SFT-wo-H. For the Qwen-specific SFT chain, \CoderQwen-SFT changes the average length from 30.5 to 30.0 tokens on aiXcoder-FIM-Evaluation and from 14.8 to 13.9 tokens on CrossCodeEval when compared with \CoderQwen-SFT-wo-H. The retained Seed-PPL Qwen SFT anchor, \CoderQwen-SFT-S, remains short at 30.5 and 13.9 tokens on the same benchmarks and intentionally serves as the starting point for the matched-data Qwen DPO variants. Situations are similar in Santacoder-fim-task.
Subsequently, the alignment stage provides the final refinement in output control.
The repetition suppression in \textbf{STG5} also contributes to output control: while its effect on \CoderSeed-DPO in aiXcoder-FIM-Evaluation is subtle (37.8 to 37.0 tokens), it is much larger in the matched Qwen DPO comparison based on \CoderQwen-SFT-S and the Seed-PPL data pipeline, where the average length drops from 29.9 tokens in \CoderQwen-DPO-wo-RS-S to 17.3 tokens in \CoderQwen-DPO-S.
On the Santacoder-fim-task benchmark, \textbf{STG5} has a clear effect for the \CoderSeed\ series, reducing the average length from 12.7 in \CoderSeed-DPO-wo-RS to 10.0 tokens in \CoderSeed-DPO.
On CrossCodeEval, our final \CoderQwen-DPO-S and \CoderSeed-DPO models produce the leanest outputs of 13.1 and 15.9 tokens, respectively.

\begin{table}[htbp]
\centering
\begin{threeparttable}
\scriptsize
\caption{Tokens of Completions Generated by Models.}
\label{tab:model_performance_gen_length}
\setlength{\tabcolsep}{2.2pt}
\begin{tabular}{l>{\centering\arraybackslash}p{1.5cm}>{\centering\arraybackslash}p{1.5cm}>{\centering\arraybackslash}p{1.5cm}>{\centering\arraybackslash}p{1.8cm}}
\toprule

\multirow{2}{*}{\textbf{Model}} & \multirow{2}{*}{\textbf{aiXcoder\tnote{*}}} & \multirow{2}{*}{\textbf{Santacoder\tnote{\dag}}} & \multirow{2}{*}{\textbf{CCE\tnote{\S}}} & \multirow{2}{*}{\textbf{C\# in CCE\tnote{\S}}} \\

 &  &  \\

\midrule

Qwen2.5-Coder-7B      & 45.4 & 11.8 & 19.1 & 18.4 \\

\midrule

\SCQwen-SFT-wo-H & \textit{30.5} & \textit{8.5} & \textit{14.8} & \textit{14.2} \\

\SCQwen-SFT & \textit{30.0} & \textit{8.4} & \textit{13.9} & \textit{13.0} \\

\SCQwen-SFT-S        & \textit{30.5} & \textit{8.4} & \textit{13.9} & \textit{13.1} \\

\SCQwen-DPO-wo-RS-S  & \textit{29.9} & \textit{8.3} & \textit{13.7} & \textit{12.9} \\

\textbf{\SCQwen-DPO-S}        & \textit{\textbf{17.3}} & \textit{\textbf{8.3}} & \textit{\textbf{13.1}} & \textit{\textbf{12.3}} \\

\midrule

Seed-Coder-8B-Base    & 80.0 & 12.4 & 23.3 & 22.1 \\

\midrule

\SCSeed-SFT-wo-H   & \textit{68.0} & \textit{12.4} & \textit{22.1} & \textit{16.9} \\

\SCSeed-SFT        & \textit{58.7} & \textit{10.9} & \textit{17.2} & \textit{22.0} \\

\SCSeed-DPO-wo-RS  & \textit{37.8} & \textit{12.7} & \textit{16.9} & \textit{17.1} \\

\textbf{\SCSeed-DPO} & \textit{37.0} & \textit{10.0} & \textit{15.9} & \textit{15.7} \\

\bottomrule
\end{tabular}
\begin{tablenotes}
\item[*] \textbf{aiXcoder}: aiXcoder-FIM-Evaluation;
\item[\dag] \textbf{Santacoder}: Santacoder-fim-task.
\item[\S] \textbf{CCE}: CrossCodeEval.
\end{tablenotes}
\end{threeparttable}
\end{table}

On CrossCodeEval overall, \CoderQwen-DPO-S (13.1 tokens) and \CoderSeed-DPO (15.9 tokens) maintain the shortest generations within the variants in their respective model families.

\section{Discussion}\label{section:discussion}

In this work, we trained \Coder\ for FIM code completion tasks by mitigating typical anti-patterns derived from common training-corpus masking strategies, as well as erroneous context repetition in compact code LLM inference.
In this section, we discuss the data-centric and practical implications of our study.

\subsection{Lessons from Data-Centric LLM Training}\label{section:data-centric-lessons}

In the current era where LLMs are increasingly used in software engineering, our work offers several data-centric lessons for training compact FIM models.

First, while the base model remains important, effectively adapting a FIM-capable code model also requires software-engineering knowledge. Our data construction process encodes common completion expectations into the training corpus, so the model is exposed to targets that better match how developers invoke inline completion.

Second, data quality should be considered at the level of constructed FIM instances, not only at the level of raw source files. Naive masking can create completion targets that are syntactically valid but misaligned with developer expectations. Our evidence focuses on the mitigation pipeline as a whole: after applying the optimized masking strategies and alignment-stage components, the trained models improve precision metrics and reduce erroneous code-echo behavior in the evaluated settings.

Finally, the AP-to-STG mapping is not a strict one-to-one taxonomy. Anti-patterns with a shared construction cause may be mitigated by one strategy, and alignment-stage strategies may provide broader signals beyond a single symptom. We therefore treat the mapping as a primary mitigation guide and validate the full design through staged ablations.
For preference alignment, our chosen responses remain ground-truth anchored; incorporating execution-validated or self-distilled valid alternatives is a promising extension, but it is outside the current pipeline.

\subsection{Practical Constraints and Economic Considerations}

For practitioners, the target deployment scenario of this work is latency-sensitive FIM inline completion. Enterprises may prefer compact open-source backbones for privacy, cost, and customization reasons, and such systems operate under tight response-time and output-length budgets. In this setting, precision alone is not sufficient: overly long completions and erroneous context echoes can increase decoding cost and force developers to inspect or delete redundant code.

Post-processing remains a useful engineering safeguard, especially in production IDE pipelines. However, post-processing can only handle an output after it has already been generated. The training-side goal of \Coder\ is therefore complementary: it reduces the model's tendency to produce erroneous echoes or unnecessarily long completions before downstream cleanup is applied. Practitioners should choose the balance between model training, decoding constraints, and post-processing according to their own latency, cost, and usability requirements.

\subsection{Potential Generalization Analysis}\label{section:potential-generalization-analysis}

Notably, C\# is absent from the post-training corpora, while the C\# subset of CrossCodeEval provides this unseen-language evaluation view with $N=1{,}768$ instances.
Nonetheless, as evidenced by the final results, the proposed training method exhibited a clear positive impact on the unseen C\# language. Specifically,
in \textbf{Table~\ref{tab:model_performance_compare_cceval}}, Seed-Coder-8B-Base records 61.9 EM and 89.1 ES on C\#, while \CoderSeed-DPO records 66.6 EM and 91.1 ES. Qwen2.5-Coder-7B records 61.5 EM and 87.8 ES, while \CoderQwen-DPO-S records 64.4 EM and 90.1 ES. \textbf{Table~\ref{tab:ablation_significance_precision}} complements these values with paired EM checks on the same C\# subset: the final DPO improvements are positive and significant for both model families, including \CoderSeed-DPO-wo-RS to \CoderSeed-DPO (+1.2 EM, $p=0.039$) and \CoderQwen-DPO-wo-RS-S to \CoderQwen-DPO-S (+5.1 EM, $p<0.001$). In \textbf{Table~\ref{tab:model_performance_rep_rate}}, the final DPO rows report 0.00\% prefix and suffix code-echo on this subset. \textbf{Table~\ref{tab:model_performance_gen_length}} reports C\# generation lengths of 12.3 tokens for \CoderQwen-DPO-S and 15.7 tokens for \CoderSeed-DPO.

These results are highly consistent, leading us to believe that, perhaps due to some inherent constraints and shared characteristics of programming languages, the anti-patterns identified by \Coder\ and the corresponding mitigation strategies possess a certain potential for generalization. However, a deeper investigation into the underlying mechanisms is beyond the scope of this paper and is left for future work.

\section{Threats to Validity}\label{section:validity}
We discuss the threats to the validity of our work.

\paragraph*{\uline{Limited Programming Languages.}} \
Due to limitations in manpower and computing resources, this study only adopted a limited set of programming languages for thematic analysis, training corpus preparation and final evaluation. This inevitably compromised the generalizability of our research. Nevertheless, the experimental results show that the identified APs and corresponding mitigation strategies remain effective for JavaScript, TypeScript and C++, none of which were included in the thematic analysis. More notably, C\# — a language absent from both the thematic analysis and training corpus — also achieved substantial and consistent performance improvements with the \Coder\ method. These findings strongly demonstrate that the \Coder\ method possesses fair generalizability.

\paragraph*{\uline{Limited Base Models.}} \
Because the evaluation covers two compact 7B/8B-scale code LLM backbones, Seed-Coder-8B-Base and Qwen2.5-Coder-7B, models at substantially smaller or larger scales, or from different architectures and families such as CodeLlama, may exhibit different behavior. The two compact base families reduce single-backbone dependence, but they do not establish size-invariant or family-invariant conclusions.

\paragraph*{\uline{Limited Benchmarks.}} \
The benchmark suite cannot cover the full diversity of real IDE FIM completion tasks, and Santacoder-fim-task becomes less discriminative once strong FIM models reach the high-90\% Pass@1 range. Accordingly, Santacoder-fim-task is interpreted mainly as a short in-file completion and stopping-behavior benchmark, while aiXcoder-FIM-Evaluation and CrossCodeEval provide complementary text-level and cross-file evidence; Santacoder-fim-task Pass@1 is also excluded from the paired significance analysis in \textbf{Table~\ref{tab:ablation_significance_precision}}.

\paragraph*{\uline{Potential Data Leakage.}} \
If benchmark data or same-repository content were present in the post-training corpus, reported performance could be inflated. While possible benchmark exposure in the unknown pretraining corpora of third-party base model is inevitable, we control this risk by removing repositories for benchmarks with identifiable provenance, together with excluded-by-construction handling for Santacoder-fim-task. These treatments are able to reduce contamination risk in the constructed post-training data.

\paragraph*{\uline{Partial Model-Matched PPL Filtering.}} \
Perplexity-based filtering depends on the reference model~\citep{ankner2024perplexed, wu2025mitigating}, and the Qwen-specific rerun covers only SFT-stage variants rather than a full two-stage Qwen-filtered pipeline or multilingual filtering study. Accordingly, the results distinguish suffix-free Qwen SFT rows from the Seed-PPL \texttt{-S} matched-data route; full model-matched filtering validation would require a separate filtering, training, and cross-language evaluation matrix. Given the obvious gains of \Coder\ at present, this may not be ROI efficient.

\paragraph*{\uline{Potentially Suboptimal Data Mixture.}} \
The STG-based data proportions are operational outcomes of the construction and filtering pipeline rather than a tuned mixture. Consequently, alternative STG5/STG6 ratios may affect repetition suppression, boundary control, and over-conservatism. The study reports the mixture used and avoids claiming optimality; systematic mixture validation would require a separate retraining and evaluation matrix across data mixtures, model families, training stages, and benchmarks, which is not feasible in practice.

\paragraph*{\uline{Potentially Suboptimal Hyperparameter Settings.}} \
The experiments do not include a systematic hyperparameter search, so the reported absolute performance and stage-wise gains may depend on settings that are not globally optimal. We apply official or widely adopted configurations so as to provide a stable engineering baseline for comparison. However, broader exploration would require substantial additional training across model families, stages, and benchmarks, which may not be feasible in practice.

\paragraph*{\uline{Partially Proxy-based Metrics.}} \
EM, ES, Pass@1, Code-echo Ratio, and Mean Generation Length are proxies rather than direct measures of developer experience, maintainability, or real IDE usability costs. Interpreting accuracy metrics together with experience-driven metrics reduces single-metric bias, while human engineer assessment and richer subjective evaluation protocols remain future work.

\paragraph*{\uline{Lack of Test-backed Preference Pair Validation.}} \
STG6 lacks test-backed preference-pair validation, i.e., the canonical middle span offers a scalable chosen response but may miss reasonable alternatives, and textual filtering of sampled rejected outputs cannot prove that every retained negative is semantically wrong. Given that large-scale FIM training spans often lack executable environments or test oracles, ground-truth anchoring and textual filtering provide a reproducible construction strategy, with execution-validated or self-distilled chosen responses left as future exploration.

\section{Conclusion}\label{section:conclusion}

This study examines data-construction issues in training FIM code completion models, including anti-patterns in common masking strategies and erroneous context repetition in compact code LLM inference. We proposed optimized masking strategies and alignment-stage components to improve the realism of training corpora and better align FIM training data with human completion expectations.

By leveraging SFT and DPO, we finetuned the Seed-Coder-8B-Base and Qwen2.5-Coder-7B models. Despite starting from compact base models under 10 billion parameters, the resulting models achieve strong performance among the evaluated FIM baselines. These results suggest that model capability is not solely determined by parameter count, but also depends on high-quality, human-aligned training data.

Moving forward, we encourage further exploration into subjective human evaluations and cost-effective deployment strategies to better bridge the gap between model performance and real-world developer needs. This study not only advances the state of FIM code completion but also provides a replicable framework for data-centric LLM training in software engineering applications.

\begin{acknowledgements}
The paper is jointly supported by both the China Telecom Research Institute and the Science and Technology Development Program of Two Districts in Xinjiang Province, China.
\end{acknowledgements}

\section*{Declarations}
\paragraph*{\upshape \textbf{Funding}} \quad
The research leading to these results received funding from the Science and Technology Development Program of Two Districts in Xinjiang Province, China under Grant No. 2024LQ03004.
\paragraph*{\upshape \textbf{Ethical approval}} \quad
Not applicable.
\paragraph*{\upshape \textbf{Informed consent}} \quad
Not applicable.
\paragraph*{\upshape \textbf{Author Contributions}} \quad
Dongjun Yu: Conceptualization, Methodology, Formal analysis, Writing -- original draft and Visualization;
Xiao Yan: Conceptualization, Methodology, Formal analysis and Writing -- original draft;
Zhenrui Li: Conceptualization, Methodology, Investigation and Writing -- original draft;
Jipeng Xiao: Visualization and Writing -- review \& editing;
Haochuan He: Writing -- review \& editing;
Yongda Yu: Methodology;
Hao Zhang: Writing -- review \& editing;
Guoping Rong: Writing -- review \& editing, Funding acquisition, Resources and Supervision;
Xiaobo Huang: Methodology, Investigation and Writing -- review \& editing.

\paragraph*{\upshape \textbf{Data Availability Statement}}  \quad
Due to corporate compliance and data privacy regulations, the source code and datasets cannot be made publicly available.
\paragraph*{\upshape \textbf{Conflict of Interest}}  \quad
The authors have no competing interests to declare that are relevant to the content of this article.
\paragraph*{\upshape \textbf{Clinical Trial Number in the manuscript}}  \quad
Not applicable.

\bibliographystyle{spbasic}
\bibliography{main}

\end{document}